\documentclass[prd,superscriptaddress,nofootinbib,preprintnumbers,amsmath,amssymb,10pt,twocolumn,dvipsnames,floatfix]{revtex4-2}
\usepackage[utf8]{inputenc}
\usepackage{amsfonts}
\usepackage{amsmath}
\usepackage{mathrsfs}
\usepackage{physics}
\usepackage{footnote}
\usepackage{scrextend}
\usepackage{leftidx}
\usepackage{soul}
\usepackage{braket}
\usepackage{makecell}
\usepackage{multirow}
\usepackage{array}
\usepackage{orcidlink} 
\usepackage{bm}
\usepackage[lofdepth,lotdepth,caption=false,justification=RaggedRight]{subfig}
\usepackage{slashed}
\usepackage{siunitx}
\usepackage{mhchem}
\usepackage[capitalize]{cleveref}
\usepackage{cancel}
\usepackage{xcolor}
\definecolor{red}{rgb}{0.9, 0,0}
\definecolor{cerulean}{rgb}{0., 0.42,0.9}

\usepackage{epsfig}
\usepackage{graphicx}  
\usepackage{url}
\usepackage{color}
\usepackage{hyperref}
\hypersetup{
     colorlinks   = true,
     citecolor    = red,
	 linkcolor=blue
}
\usepackage[capitalize]{cleveref}    
  \crefname{algorithm}{Alg.}{Algs.}
  \crefname{tab}{Table}{Tables}
  \crefname{fig}{Fig.}{Figs.}
  \crefname{section}{Sec.}{Secs.}
  \crefname{appendix}{App.}{Apps.}
\graphicspath{{./figure/}{./circuits}}
\allowdisplaybreaks

\usepackage{titlesec}
\titlespacing*{\section}
{0pt}{18pt plus 4pt minus 2pt}{10pt plus 4pt minus 2pt}
\titlespacing*{\subsection}
{0pt}{12pt plus 4pt minus 2pt}{10pt plus 4pt minus 2pt}

\titleformat{\section}
  {\centering\small\bfseries\MakeUppercase} 
  {\thesection.}{1em}{}

\titleformat{\subsection}
  {\centering\normalfont\small\bfseries} 
  {\thesubsection.}{1em}{}

\newcommand{\appendixsectionformat}{%
  \titleformat{\section}
    {\centering\normalfont\bfseries} 
    {\thesection.}{1em}{}
}

\makeatletter 
    
\renewcommand\onecolumngrid{
\do@columngrid{one}{\@ne}%
\def\set@footnotewidth{\onecolumngrid}
\def\footnoterule{\kern-6pt\hrule width 1.5in\kern6pt}%
}

\makeatother   

\newcommand{\spn}[1]{{\mathrm{span}\{#1\}}}
\newcommand{\Ntr}{{\ce{^14N}}}
\newcommand{\Nobs}[1][]{{N_{\mathrm{obs}}^{#1}}}
\newcommand{\tobs}[1][]{{t_{\mathrm{obs}}^{#1}}}

\newcommand{\flabel}{{\chi}}
\newcommand{\MyTr}[1]{{\mathrm{Tr}\,\left[ {#1} \right]}}
\newcommand{\PSD}[2][]{{\mathcal{P}^{#1}_{#2}}}
\newcommand{\SPSD}[2][]{{\mathcal{S}^{#1}_{#2}}}
\newcommand{\BPSD}[2][]{{\mathcal{B}^{#1}_{#2}}}
\newcommand{\fluorescence}{{F}}
\newcommand{\navg}[1][]{{n_{\mathrm{avg}}^{#1}}}

\begin{document}


\preprint{KEK-QUP-2024-0013, TU-1233}

\title{Nuclear Spin Metrology with Nitrogen Vacancy Center in Diamond for \\ Axion Dark Matter Detection}

\author{So Chigusa\,\orcidlink{0000-0001-6005-4447}}
\affiliation{Theoretical Physics Group, Lawrence Berkeley National Laboratory, Berkeley, CA 94720, USA}
\affiliation{Berkeley Center for Theoretical Physics, Department of Physics, University of California, Berkeley, CA 94720, USA}
\affiliation{International Center for Quantum-field Measurement Systems for Studies of the Universe and Particles (QUP), High Energy Accelerator Research Organization (KEK), 1-1 Oho, Tsukuba, Ibaraki 305-0801, Japan}

\author{Masashi Hazumi}
\affiliation{International Center for Quantum-field Measurement Systems for Studies of the Universe and Particles (QUP), High Energy Accelerator Research Organization (KEK), 1-1 Oho, Tsukuba, Ibaraki 305-0801, Japan}
\affiliation{Institute of Particle and Nuclear Studies (IPNS), KEK, Tsukuba, Ibaraki 305-0801, Japan}
\affiliation{Japan Aerospace Exploration Agency (JAXA), Institute of Space and Astronautical Science (ISAS), Sagamihara, Kanagawa 252-5210, Japan}
\affiliation{Kavli Institute for the Physics and Mathematics of the Universe (Kavli IPMU, WPI), UTIAS, The University of Tokyo, Kashiwa, Chiba 277-8583, Japan}
\affiliation{The Graduate University for Advanced Studies (SOKENDAI), Miura District, Kanagawa 240-0115, Hayama, Japan}
\affiliation{Department of Physics and Center for High Energy and High Field Physics, National Central University, Zhongli District, Taoyuan City, Taiwan}

\author{Ernst David Herbschleb\,\orcidlink{0000-0001-7579-1765}}
\affiliation{Institute for Chemical Research, Kyoto University, Gokasho, Uji-city, Kyoto 611-0011, Japan}

\author{Yuichiro Matsuzaki}
\affiliation{
Department of Electrical, Electronic, and Communication Engineering, Faculty of Science and Engineering, Chuo university, 1-13-27, Kasuga, Bunkyo-ku, Tokyo 112-8551, Japan
}

\author{Norikazu Mizuochi}
\affiliation{International Center for Quantum-field Measurement Systems for Studies of the Universe and Particles (QUP), High Energy Accelerator Research Organization (KEK), 1-1 Oho, Tsukuba, Ibaraki 305-0801, Japan}
\affiliation{Institute for Chemical Research, Kyoto University, Gokasho, Uji-city, Kyoto 611-0011, Japan}
\affiliation{Center for Spintronics Research Network, Kyoto University, Uji, Kyoto 611-0011, Japan}

\author{Kazunori Nakayama}
\affiliation{International Center for Quantum-field Measurement Systems for Studies of the Universe and Particles (QUP), High Energy Accelerator Research Organization (KEK), 1-1 Oho, Tsukuba, Ibaraki 305-0801, Japan}
\affiliation{Department of Physics, Tohoku University, Sendai, Miyagi 980-8578, Japan}

\begin{abstract}
We present a method to directly detect the axion dark matter using nitrogen vacancy centers in diamonds.
In particular, we use metrology leveraging the nuclear spin of nitrogen to detect axion-nucleus couplings.
This is achieved through protocols designed for dark matter searches, which introduce a novel approach of quantum sensing techniques based on the nitrogen vacancy center.
Although the coupling strength of the magnetic fields with nuclear spins is three orders of magnitude smaller than that with electron spins for conventional magnetometry, the axion interaction strength with nuclear spins is the same order of magnitude as that with electron spins.
Furthermore, we can take advantage of the long coherence time by using the nuclear spins for the axion dark matter detection.
Our method has the potential to be sensitive to a broad frequency range $\lesssim \SI{100}{Hz}$ corresponding to the axion mass $m_a \lesssim \SI{4e-13}{eV}$.
We present the detection limit of our method for both the axion-neutron and the axion-proton couplings and discuss its significance in comparison with other proposed ideas.
We also show that the sensitivities of the NV center sensor to various spin species will open up new directions for constructing protocols that can mitigate magnetic noise effects.

\end{abstract}

\maketitle
\tableofcontents

\section{Introduction}

The existence of dark matter (DM) is one of the most important hints of new physics in particle physics.
While the relic abundance of DM in the current Universe is known through various cosmological and astrophysical observations, such as the galaxy rotation curve, weak gravitational lensing, and the cosmic microwave background (see, e.g.\ \cite{1933AcHPh...6..110Z,1936ApJ....83...23S,1970ApJ...159..379R,1970ApJ...160..811F,1973ApJ...186..467O,2004IAUS..220..439H,Clowe_2006,Bennett_2013,Planck:2018vyg}), other properties of DM remain unrevealed.
One approach to study these properties is through direct detection of DM in lab-based experiments.
Given the variety of DM candidates that can explain the relic abundance, numerous approaches have been taken to investigate different types of DM interactions with standard model (SM) particles (see \cite{Workman:2022ynf,Billard:2021uyg,Pato:2010zk} for a review).

Among various DM candidates, the axion stands out as a promising candidate, motivated by several contexts.
The term ``axion'' can refer to the QCD axions, such as those proposed in \cite{PhysRevLett.43.103,SHIFMAN1980493,DINE1981199,Zhitnitsky:1980tq}, which are introduced to solve the strong CP problem~\cite{Peccei:1977hh,Weinberg:1977ma,Wilczek:1977pj}, or the axion-like particles, which represent a broader set of pseudoscalar particles often predicted in low-energy effective theories emerging from the string theory~\cite{WITTEN1984351,Svrcek:2006yi,Conlon:2006tq,Choi:2009jt,Arvanitaki:2009fg,Acharya:2010zx,Cicoli:2012sz,Halverson:2017deq,Demirtas:2018akl}.
Generally, the axion interacts with SM gauge bosons and fermions, each interaction controlled by a model-dependent coupling constant.
Therefore, developing various strategies to investigate different couplings is essential to discover axions and differentiate between axion models.

In this paper, we explore the nitrogen-vacancy (NV) center in diamond, a well-studied multimodal quantum sensing device, as an apparatus for axion DM searches.
Unlike a previous study \cite{Chigusa:2023hms}, where some of the authors used NV center metrology based on electron spins to detect DM signals, we utilize the nuclear spin of the $\Ntr$ atom to search for signals induced by axion-nucleus couplings.
This approach aims to constrain the axion-neutron coupling $g_{ann}$ and the axion-proton coupling $g_{app}$, which are independent of the constraint on the axion-electron coupling $g_{aee}$ obtained in \cite{Chigusa:2023hms}.
Our method can be viewed as magnetometry based on nuclear spins.
Although this procedure is not well-suited for detecting ordinary magnetic fields due to their weak coupling to nuclear spins, it is crucial for axion DM searches because $g_{ann}$, $g_{app}$, and $g_{aee}$ are independent parameters.

The rest of the paper is organized as follows.
In \cref{sec:NV_center}, we review NV center metrology, starting with an overview of the NV center system (\cref{sec:NV_center_review}) and explaining the protocols used for dc (\cref{sec:dc_magnetometry}) and ac (\cref{sec:ac_magnetometry}) magnetometry.
\cref{sec:axion} reviews axion properties, where we derive the axion interaction Hamiltonian with elementary particles (\cref{sec:axion_setup}) and the $\Ntr$ spin (\cref{sec:axion_interaction}).
We discuss our detection limit estimation in \cref{sec:sensitivity} and present constraints on the axion coupling constants in \cref{sec:results}.
Finally, we provide concluding remarks in \cref{sec:conclusion}.

\section{NV center metrology}
\label{sec:NV_center}

\subsection{NV center in diamonds}
\label{sec:NV_center_review}

The NV center is a complex composed of a substitutional nitrogen and an adjacent vacancy.
Among the various possible charge states, $\mathrm{NV}^-$ is often used for quantum sensing, where two remnant electrons are localized to the position of the vacancy.
These two electrons form the orbital-singlet, spin-triplet system at the lowest energy levels.
The other possible combinations of angular momenta, which include the orbital triplet and/or the spin-singlet states, correspond to excited states.
The electron system is excited to an orbital-triplet state by injecting $\SI{532}{nm}$ green light, which can relax either directly with emitting $600$--$\SI{800}{nm}$ red light or through spin-singlet states with emitting infrared light.
Since the probability of direct relaxation depends on whether the initial state of the two-electron spin $\vec{S}$ is $\ket{S^z=0}$ or $\ket{S^z=\pm}$, we can read out the spin state information through the fluorescence measurement.
When relaxing through the spin-singlet states, the electron spin usually ends up in the lowest energy state $\ket{S^z=0}$, thus making the whole procedure work also as laser cooling.

In addition to the electrons at the NV center, the substitutional nitrogen also possesses a (nuclear) spin degree of freedom, $\vec{I}$.
Since $\sim 99.6\%$ of the nitrogens in nature are $\Ntr$ with spin $I=1$, we focus on this isotope.
Including the hyperfine interaction between the electron and nuclear spins, the dynamics of the NV center spin system is governed by the Hamiltonian
\begin{align}
  H &= H_\parallel + H_\perp,
\end{align}
where the first (second) term corresponds to the interactions parallel to (perpendicular to) the $z$-axis, which is defined by the NV axis.\footnote{
  According to this definition of the $z$-axis, four different configurations of the NV center in the diamond lattice are effectively distinguished by choosing four different sets of local coordinates, resulting in different effective magnetic fields.
  This affects the resonance frequency of the Rabi cycle we will discuss below, thus the succeeding spin operation is effective only for a part of four configurations.
  However, note that the orientation can be aligned with a specific fabrication process~\cite{Fukui:2014,Miyazaki:2014,Michl:2014,Lesik:2014}.
}
They are given by~\cite{RevModPhys.92.015004}
\begin{align}
  H_\parallel &= \Delta_0 S^{z2} + Q_0 I^{z2} + B^z (\gamma_e S^z + \gamma_N I^z) + A_\parallel S^z I^z, \\
  H_\perp &= \gamma_e \vec{B}_\perp \cdot \vec{S}_\perp + \gamma_N \vec{B}_\perp \cdot \vec{I}_\perp + A_\perp \vec{S}_\perp \cdot \vec{I}_\perp, 
\end{align}
where $\vec{B}$ is an external magnetic field, while the subscript $_\perp$ of a vector denotes components perpendicular to the $z$-axis.
$\Delta_0 \simeq 2\pi\times \SI{2.87}{GHz}$ and $Q_0 \simeq -2\pi\times \SI{4.95}{MHz}$ are the zero-field splitting of electron spins and the nuclear quadrupole interaction parameter, respectively.
The gyromagnetic ratios for electron and nuclear spins are given respectively by $\gamma_e \simeq 2\pi\times \SI{28}{GHz/T}$, $\gamma_N = 2\pi\times \SI{3.08}{MHz/T}$~\cite{2004960}.
The size of the hyperfine interaction is measured as $A_\parallel \simeq -2\pi\times \SI{2.16}{MHz}$ and $A_\perp = -2\pi\times \SI{2.62}{MHz}$.

\begin{figure}
  \centering
  \includegraphics[width=\hsize]{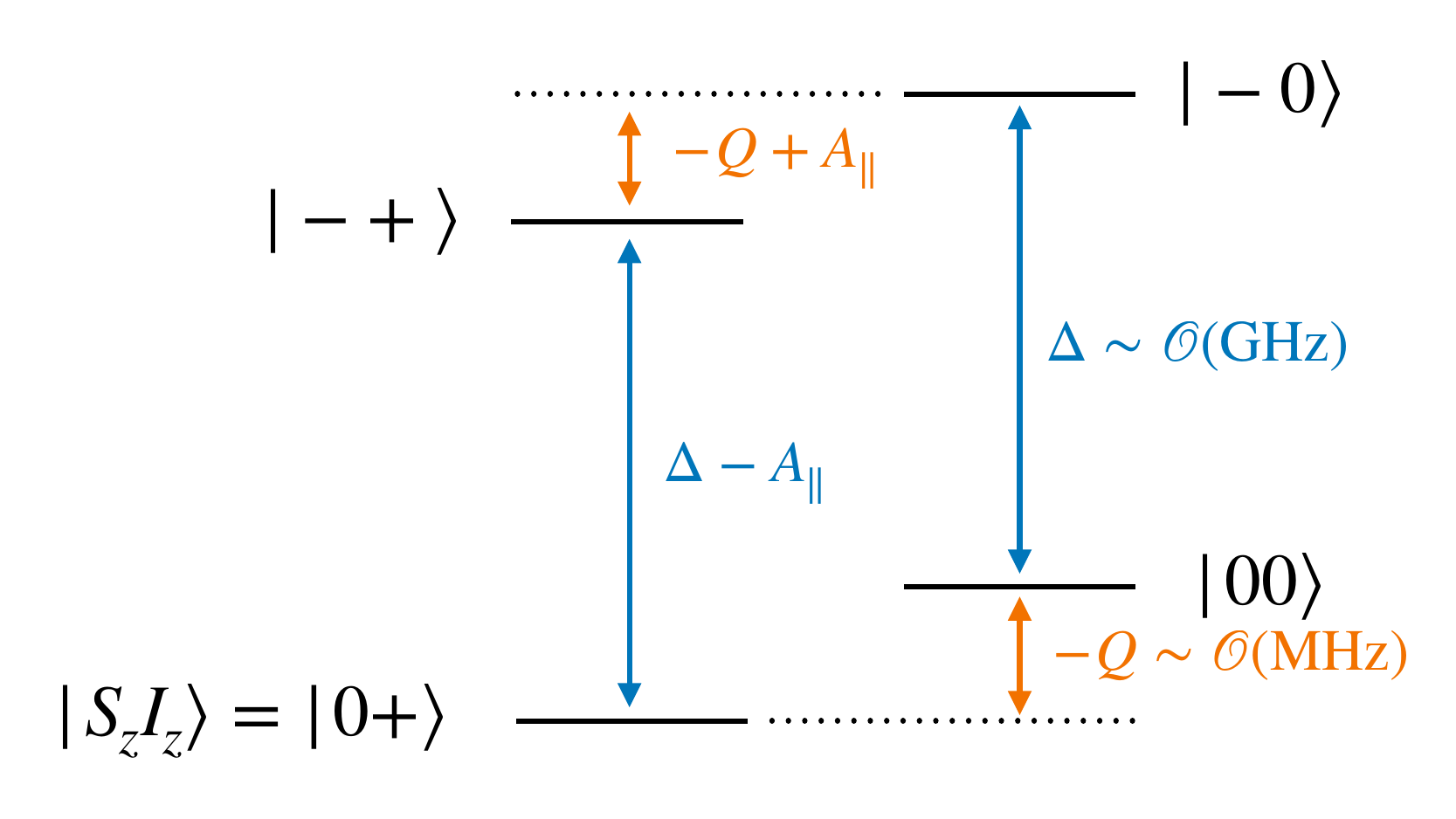}
  \caption{Energy levels of the two-qubit subsystem, with $\Delta$ and $Q$ described in the text, and $A_\parallel$ the size of the hyperfine splitting.}
  \label{fig:energy-levels}
\end{figure}

The quantum state of electron and nuclear spins can be manipulated by the Rabi cycle.
To see this in more detail, let us first pick up two of the electron spin states, say $\ket{S^z=-,0}$, and two of the nuclear spin states, say $\ket{I^z=0,+}$, to constrain ourselves to an effective two-qubit subsystem spanned by $\{ \ket{S^z I^z} = \ket{-0}, \ket{-+}, \ket{00}, \ket{0+} \}$.
For later convenience, we assume the decomposition $\vec{B} = B_0 \hat{z} + \vec{B}_\perp \cos \omega t$ with $\hat{z}$ the unit vector along the $z$-axis. 
If we treat $H_\perp$, or $\vec{B}_\perp$ and $A_\perp$, as a perturbation, four states labelled by $\ket{S^z I^z}$ are energy eigenstates, whose energy levels are shown in \cref{fig:energy-levels} with $\Delta \equiv \Delta_0 - \gamma_e B_0$ and $Q \equiv Q_0 + \gamma_N B_0$.
The effect of the oscillating transverse magnetic field is described by the effective Hamiltonian
\begin{align}
  H_{\mathrm{eff}}(t) =& \left[
    \frac{1}{\sqrt{2}} \gamma_e \left(
      \vec{B}_\perp \cdot \vec{\sigma}_e
    \right) + \frac{1}{\sqrt{2}} \gamma_N \left(
      \vec{B}_\perp \cdot \vec{\sigma}_n
    \right)
  \right] \notag \\
  &\times \cos\omega t,
\end{align}
up to constant terms, where $\vec{\sigma}_e$ and $\vec{\sigma}_n$ are Pauli matrices $\vec{\sigma}$ acting on the corresponding qubit $\ket{e} \in \spn{\ket{S^z = -,0}}$ and $\ket{N} \in \spn{ \ket{I^z = 0, +} }$, respectively.
In the vector space with the basis choice of $\{ \ket{-0}, \ket{-+}, \ket{00}, \ket{0+} \}$, we obtain the matrix representation
\begin{align}
  H_{\mathrm{eff}}(t) = \frac{1}{\sqrt{2}} \begin{pmatrix}
    0 & \gamma_N B^{-} & \gamma_e B^{-} & 0 \\
    \gamma_N B^{+} & 0 & 0 & \gamma_e B^{-} \\
    \gamma_e B^{+} & 0 & 0 & \gamma_N B^{-} \\
    0 & \gamma_e B^{+} & \gamma_N B^{+} & 0
  \end{pmatrix} \cos\omega t,
\end{align}
with $B^{\pm} \equiv B^x \pm i B^y$.

It is convenient to work in the interaction picture with $H_\parallel$ treated as the unperturbed Hamiltonian.
The effective Hamiltonian is then given by $\tilde{H}_{\mathrm{eff}}(t) \equiv e^{i H_\parallel t} H_{\mathrm{eff}}(t) e^{-i H_\parallel t}$ or
\begin{widetext}
\begin{align}
  \tilde{H}_{\mathrm{eff}}(t) = \frac{1}{\sqrt{2}} \begin{pmatrix}
    0 & \gamma_N B^{-} e^{i(-Q+A_\parallel)t} & \gamma_e B^{-} e^{i\Delta t} & 0 \\
    \gamma_N B^{+} e^{-i(-Q+A_\parallel)t} & 0 & 0 & \gamma_e B^{-} e^{i(\Delta - A_\parallel)t} \\
\gamma_e B^{+} e^{-i\Delta t} & 0 & 0 & \gamma_N B^{-} e^{-iQt} \\
    0 & \gamma_e B^{+} e^{-i(\Delta - A_\parallel)t} & \gamma_N B^{+} e^{iQt} & 0
  \end{pmatrix} \cos\omega t.
\end{align}
\end{widetext}
By noting that the fast oscillation terms in the above expression can be neglected, it becomes clear the oscillating magnetic field drives transformation between two energy levels whose energy gap is close to the oscillation frequency $\omega$.
For example, if we start from $\ket{\psi(t=0)} = \ket{0+}$ and choose $\omega = \Delta-A_\parallel$, which is typically in the microwave frequency range, the dynamics of the quantum state are given by
\begin{align}
  \ket{\psi(t)} = \exp \left(
    \frac{i}{\sqrt{2}} \gamma_e \vec{B}_\perp\cdot\vec{\sigma} t
  \right) \begin{pmatrix}
    \ket{-+} \\
    \ket{0+}
  \end{pmatrix}.
  \label{eq:Rperp_e}
\end{align}
Thus, a manipulation of the electron spin state that only affects states with $\ket{I^z=+}$ is possible.
Similar dynamics controlled on $\ket{I^z=0}$ are realized with the choice $\omega = \Delta$.
If we choose $\omega = -Q$ instead, which is typically in the radio frequency range, the dynamics are expressed as
\begin{align}
  \ket{\psi(t)} = \exp \left(
    \frac{i}{\sqrt{2}} \gamma_N \vec{B}_\perp\cdot\vec{\sigma} t
  \right) \begin{pmatrix}
    \ket{00} \\
    \ket{0+}
  \end{pmatrix}.
  \label{eq:Rperp_n}
\end{align}
Thus, a manipulation of the nuclear spin state that only affects states with $\ket{S^z=0}$ is possible also.
Similar dynamics controlled on the $\ket{S^z = -}$ state are realized with the choice $\omega = -Q+A_\parallel$.
Note that energy gaps relevant to the neglected five energy eigenstates composed of $\ket{S^z=+}$ and/or $\ket{I^z=-}$ take different values under non-zero $B_0$, so we can stick to the effective two-qubit system of the total Hilbert space by simply restricting ourselves to the four relevant frequencies, i.e.\ $\Delta - A_\parallel$, $\Delta$, $-Q$, and $-Q+A_\parallel$.

The dynamics described in \cref{eq:Rperp_e} (\cref{eq:Rperp_n}) with $\vec{B}_\perp \propto \hat{x}$ and $\hat{y}$ represent the (controlled-)$R_x$ and $R_y$ gates acting on the qubit $\ket{e}$ ($\ket{N}$), respectively, with a tunable rotation angle $\theta = \sqrt{2} \gamma_e B_\perp t$ ($\theta = \sqrt{2} \gamma_n B_\perp t$).
As is well known, by combining $R_x$ and $R_y$ gate operations one can construct an arbitrary $SU(2)$ operation acting on the target qubit.
By also noting that the controlled-$R_x$ gate with $\theta = \pi$ (or simply $\pi_x$) works as the CNOT gate up to a global phase factor, an arbitrary $SU(4)$ operation on the two-qubit system can in principle be implemented~\cite{PhysRevA.69.010301}.
Finally, projection measurement of the nuclear spin qubit $\ket{N}$ can be performed by combining the CNOT gate acting on the electron spin and the preceding fluorescence measurement~\cite{doi:10.1126/science.1189075}.
Physically, this final CNOT gate operation is done with a $\pi$-pulse with frequency $\Delta$ so that the electron spin state is excited only when the nuclear spin state is $\left|0\right>$.
In the qubit picture, signal strength of the fluorescence measurement is characterized by
\begin{align}
  \fluorescence \equiv \frac{1}{2} \Braket{\varphi | \sigma^z | \varphi},
\end{align}
where $\ket{\varphi}$ is the qubit state of the electron or nuclear spin depending on the setup.

Thanks to the available spin state manipulation and measurement described so far, the NV center works as a multimodal quantum sensor~\cite{RevModPhys.92.015004}.
Technologies to realize the high nuclear-spin polarization via CNOT gates, e.g. near $99\,\%$ was demonstrated for single nuclear spins \cite{doi:10.1126/sciadv.abg9204}, make it possible for the nuclear spins to be properly initialized.
Additionally, we can operate with either a single NV center or an ensemble of NV centers \cite{Barry:2023hon,10.1063/1.5047078}.
In this paper, we focus on the latter choice with which a large number of NV centers, $N\gg 1$, helps improve the sensitivity by accumulation of large statistics.

\subsection{Dc magnetometry}
\label{sec:dc_magnetometry}

Now, we describe the so-called Ramsey sequence \cite{PhysRev.78.695} used for dc magnetometry.
Throughout this and the next subsections, we focus on the evolution of a qubit state $\ket{\varphi(t)}$, which can be either the electron or nuclear spin state.
In the matrix representation, we use the basis $\{ \ket{S^z} = \ket{-}, \ket{0} \}$ for the electron spin and $\{ \ket{I^z} = \ket{0}, \ket{+} \}$ for the nuclear spin.

The Ramsey sequence is sensitive to a dc-like magnetic field $B_s(t) \hat{z}$ along the $z$-axis.\footnote{
The magnetic field in the $xy$-plane can be neglected as long as its oscillation frequency is far from the energy gap of the qubit system.
See the calculation of the Rabi cycle.
}
Let $\tilde{H}_{\mathrm{int}}(t)$ be the corresponding interaction Hamiltonian in the interaction picture defined as
\begin{align}
  \tilde{H}_{\mathrm{int}}(t) = \frac{1}{2} \gamma B_s(t) \sigma^z,
\end{align}
where $\gamma = \gamma_e$ or $\gamma_N$ is the suitable choice of the gyromagnetic ratio.
Starting from the lower level $\ket{\varphi(0)} = (0,1)^\intercal$, the qubit state evolution under the Ramsey sequence is given by\footnote{
In this expression and the later discussion, we neglect the time spent on gate operations for simplicity.
It is a reasonable approximation when the amplitude of the magnetic pulse used for spin operations is large enough as can be seen from \cref{eq:Rperp_n}.
}
\begin{align}
  \ket{\varphi(\tau)} = R_x^{\pi/2} \exp \left(
    -i \int_0^\tau dt\, \tilde{H}_{\mathrm{int}} (t)
  \right) R_y^{\pi/2} \begin{pmatrix}
    0 \\
    1
  \end{pmatrix},
  \label{eq:qs_dc}
\end{align}
where $\tau$ is the time duration of free precession, while $R_\alpha^\theta$ $(\alpha=x,y)$ denotes the corresponding $R_\alpha$ gate operation represented in matrices as
\begin{align}
  R_x^\theta &= \begin{pmatrix}
    \cos\frac{\theta}{2} & -i \sin\frac{\theta}{2} \\
    -i \sin\frac{\theta}{2} & \cos\frac{\theta}{2}
  \end{pmatrix}, \\
  R_y^\theta &= \begin{pmatrix}
    \cos\frac{\theta}{2} & -\sin\frac{\theta}{2} \\
    \sin\frac{\theta}{2} & \cos\frac{\theta}{2}
  \end{pmatrix}.
\end{align}
If the signal magnetic field oscillates slowly as $B_s(t) = B_s^0 \cos(\epsilon t + \phi)$, the signal strength of the fluorescence measurement for the state $\ket{\varphi(t)}$ is explicitly calculated as
\begin{align}
  \fluorescence \simeq \frac{\gamma B_s^0}{2 \epsilon} \left[
    \sin(\epsilon \tau + \phi) - \sin \phi
  \right],
  \label{eq:S_DC}
\end{align}
under the assumption of $\fluorescence \ll 1$.
It takes a constant value $\fluorescence\to \gamma B_s^0 \tau \cos\phi / 2$ at $\epsilon\to 0$, while the cancellation of fast oscillations leads to an asymptotic behavior $\fluorescence \propto \epsilon^{-1}$ when $\epsilon \tau \gtrsim 1$.
Thus, this approach is effective for a dc-like signal with an angular frequency $\epsilon \ll 1/\tau$.

\begin{figure}[htp]
  \centering
  \includegraphics*[width=\hsize]{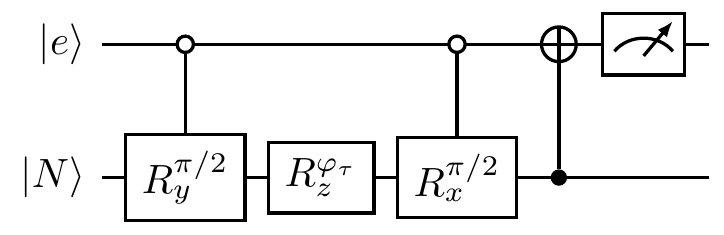}
  \caption{The protocol of dc magnetometry using nuclear spins.}
  \label{fig:protocol_DC}
\end{figure}

When considering an ordinary magnetic field, electron spins are more useful than nuclear spins to obtain a sizable effect within a fixed time duration $\tau$ due to the hierarchy $\gamma_e \gg \gamma_N$.
However, this is not the case for the axion dark matter detection, because, as we will see below, the axion interaction strength with electron and nuclear spins have completely different, and model dependent, relationships.
Therefore, it is motivated to explore nuclear-spin-based dc magnetometry as a complementary probe to the one based on electron spins~\cite{Chigusa:2023hms}.
In \cref{fig:protocol_DC}, we show the protocol for dc magnetometry using nuclear spins.
Both of the qubits should be initialized to $(0, 1)^\intercal$ through laser cooling and an appropriate operation of the CNOT gates before starting the protocol, preparing the $\ket{00}$ state.
$\varphi_\tau \simeq \fluorescence$ is the relative phase factor generated during the free precession.

We have not taken account of the effects of relaxation in the above expression.
There are two different relaxation time scales for each spin species, the longitudinal relaxation time $T_1$ and the transverse relaxation, or dephasing, time $T_2^*$.
$T_1$ characterizes the spin flip associated with the energy transfer to or from the environment, which takes $T_{1e} \sim \SI{6}{ms}$~\cite{PhysRevLett.108.197601, PhysRevLett.112.147602} and $T_{1N} \sim \SI{4}{min}$~\cite{doi:10.1126/science.aam8697} for the electron and nuclear spins, respectively, at room temperature.
The dominant source of the transverse relaxation, on the other hand, is dephasing of spins due to the inhomogeneous dc magnetic field caused by, e.g.\ nuclear spins or lattice defects.
$T_{2e}^{*} \sim \SI{1}{\mu s}$~\cite{RevModPhys.92.015004, Mizuochi:2009} and $T_{2N}^{*} \sim \SI{7.25}{ms}$~\cite{Waldherr2012}\footnote{
Although the measured value of $T_{2N}^{*} \sim \SI{7.25}{ms}$ is for a single NV center, we use this value as a reasonable estimate of $T_{2N}^{*}$ for an ensemble of NV centers, since for a $\sim \SI{1}{ppm}$ concentration of NV centers, $T_{1e}$ is still in the order of milliseconds~\cite{Jarmola2015}.
} are measured at room temperature.
The large hierarchy $T_{2N}^{*} / T_{2e}^{*} \sim 10^4$ is consistent with the large hierarchy of interaction strengths $\gamma_e / \gamma_N \sim 10^4$.
It should be noted that $T_{2N}^{*}$ is naturally bounded by $T_{1e}$ since random electron spin flips induce dephasing of nuclear spins through the hyperfine interaction.
On the other hand, $T_{2e}^{*}$ does not necessarily limit $T_{2N}^{*}$.
For example, the dephasing time scale of the quantum states $(\ket{0+}+\ket{00})/\sqrt{2}$ would be $T_{2N}^{*}$.
Our protocol for nuclear spin dc magnetometry corresponds to this case.

Taking into account the effects of relaxation, the signal strength in \cref{eq:S_DC} is rescaled as $\fluorescence\to \fluorescence e^{-\tau/T_{2N}^{*}}$.
Accordingly, the optimistic choice of $\tau$ to maximize the sensitivity turns out to be $\tau\sim T_{2N}^{*}/2$~\cite{Herbschleb:2019}.
We will use this choice for later analysis.

\subsection{Ac magnetometry}
\label{sec:ac_magnetometry}

\begin{figure}[htp]
  \centering
  \includegraphics*[width=\hsize]{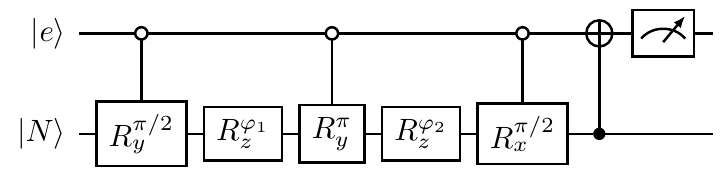}
  \caption{
    The protocol of ac magnetometry using nuclear spins.
    $\varphi_1$ and $\varphi_2$ represent the phase factors acquired in the first half ($0<t<\tau/2$) and the second half ($\tau/2<t<\tau$) of the free precession time, respectively.
  }
  \label{fig:protocol_AC}
\end{figure}

As we have seen thus far, the Ramsey sequence is not effective for an ac magnetic field with an angular frequency $\epsilon \gtrsim 1/\tau$.
To realize another approach sensitive to such high-frequency signals, we can make use of the Hahn-echo sequence~\cite{PhysRev.80.580} or dynamical decoupling sequences~\cite{PhysRev.94.630, meiboom1958modified} in more general context.
The time evolution of a qubit state under the Hahn echo sequence is described by
\begin{widetext}
\begin{align}
  \ket{\varphi(\tau)} = R_x^{\pi/2} \exp \left(
    -i \int_{\tau/2}^\tau dt\, \tilde{H}_{\mathrm{int}} (t)
  \right) R_y^{\pi} \exp \left(
    -i \int_0^{\tau/2} dt\, \tilde{H}_{\mathrm{int}} (t)
  \right) R_y^{\pi/2} \begin{pmatrix}
    0 \\
    1
  \end{pmatrix},
  \label{eq:qs_ac}
\end{align}
\end{widetext}
where the only difference from the Ramsey sequence is the $\pi_y$ operation at the middle of the free precession.
This operation reverses the effect from the signal magnetic field and achieves constructive interference of the oscillating signal effect before and after the $\pi_y$ operation when the angular frequency is $\sim 2\pi/\tau$.
The signal strength is explicitly calculated as follows:
\begin{align}
  \fluorescence = \frac{2\gamma B_s^0}{\epsilon} \sin^2 \frac{\epsilon \tau}{4} \sin \left(
    \frac{\epsilon \tau}{2} + \phi
  \right),
  \label{eq:S_AC}
\end{align}
which is suppressed in both the dc limit $\epsilon\to 0$ as $S\propto \epsilon$ and the high frequency limit $\epsilon\tau\gg 1$ as $S\propto\epsilon^{-1}$.
On the other hand, it peaks at $\epsilon = 2\pi/\tau$ with a peak height $|S|=(\gamma B_s^0 \sin\phi) / \pi$ and a peak width $\Delta\epsilon \sim 1/\tau$.
This calculation indicates a narrow-band sensitivity of the Hahn echo sequence around $\epsilon \sim 2\pi/\tau$.
\cref{fig:protocol_AC} shows the protocol of the Hahn echo sequence applied to the nuclear spin qubit $\ket{N}$.

In the Hahn echo sequence, the relevant transverse relaxation time $T_2$ tends to be longer than $T_2^{*}$ for the Ramsey sequence because any dc magnetic noise effect cancels out due to the $\pi_y$ operation.
The dominant contribution to $T_2$ is the decoherence effect caused by dipole-dipole self-interaction among spins.
The observed value of $T_{2e} \sim \SI{100}{\mu s}$~\cite{Wolf:2015} at room temperature shows a two orders of magnitude enhancement compared with $T_{2e}^{*} \sim \SI{1}{\mu s}$.
Conversely, the observed value of $T_{2N} \sim \SI{10}{m s}$ at room temperature is comparable to $T_{2N}^{*}$ because both of them are limited by the single parameter $T_{1e}$.
However, both $T_1$ and $T_2$ can be further extended with more sophisticated setups.
One possibility is to consider a cryogenic environment; for example, $T_{1e} \sim 100$~s is reported at $\lesssim \SI{50}{K}$~\cite{PhysRevLett.108.197601}, where $T_{2N} \sim \SI{100}{s}, T_{2N}^{*} \sim \SI{1}{s}$ seem to be a reasonable assumption.
Another possibility is to perform a dynamic decoupling (DD) sequence with a large number ($N_\pi$) of $\pi_y$ pulses during the free precession, which also contributes to making the coherence time longer.
In this case, we obtain the signal strength as
\begin{align}
  \fluorescence =& \frac{\gamma B_s^0}{\epsilon} \sin\frac{\epsilon\tau}{2} \sin\left(
    \frac{\epsilon\tau}{2} + \phi
  \right) \tan\frac{\epsilon\tau}{2(N_\pi + 1)},
  \label{eq:S_DD}
\end{align}
which recovers \cref{eq:S_AC} for $N_\pi = 1$.
Under a DD sequence with $N_\pi \gg 1$, one expects not only a longer $T_{2N}$, but also a sensitivity peak located at a higher angular frequency $N_\pi / \tau$.

Similar to the dc magnetometry case, the relaxation effect rescales the signal strength \cref{eq:S_AC} as $\fluorescence\to \fluorescence e^{-\tau/T_{2N}}$, and the optimal choice of $\tau$ turns out to be $\tau = T_{2N}/2$.

\section{Axion dark matter}
\label{sec:axion}

\subsection{Setup}
\label{sec:axion_setup}

The axion can account for the total relic abundance of DM through mechanisms such as the misalignment mechanism~\cite{PRESKILL1983127,ABBOTT1983133,DINE1983137} or production from topological defects (see Refs.~\cite{Kawasaki:2013ae,Marsh:2015xka,DiLuzio:2020wdo} for reviews).
A wide range of the axion mass $m_a$ could be consistent with the DM relic abundance, as small as $m_a\sim \SI{e-22}{eV}$, below which the model is inconsistent with the existence of DM-dominated dwarf galaxies~\cite{Hu:2000ke}, and as large as $m_a \gtrsim O(1)\,\si{eV}$, where cosmological and astrophysical constraints on the axion DM tend to become severe (see, for example, plots in \cite{OHare:2020}).
The axion is described by a classical field experiencing coherent oscillation
\begin{align}
  a(t, \vec{x}) = a_0 \sin(m_a t + m_a \vec{v}_a \cdot \vec{x} + \phi),
  \label{eq:coherent_oscillation}
\end{align}
where $\vec{v}_a$ is the axion velocity.
Considering the energy density stored in the coherent oscillation, a relationship $\rho_a = (m_a a_0)^2 / 2$ holds, where $\rho_a \sim \SI{0.4}{GeV/cm^3}$~\cite{ParticleDataGroup:2022pth} is the local energy density of DM.
Note that the velocity $\vec{v}_a$ and the oscillation phase $\phi$ are constant only within the de Broglie wavelength $\sim (m_a v_a)^{-1}$.
Thus, $\vec{v}_a$ and $\phi$ observed at the laboratory vary with the time scale of $\tau_a \sim (m_a v_a^2)^{-1}$, which is called the coherence time.
Assuming that the axion DM halo is virialized, its typical velocity is estimated as $v_a \sim 10^{-3}$, leading to the order estimate of
\begin{align}
  \tau_a \sim \SI{6.6}{s} \left(
    \frac{\SI{e-10}{eV}}{m_a}
  \right).
\end{align}

Due to its pseudoscalar nature, the axion generally interacts with the SM fermions $\psi_\flabel$ of the form
\begin{align}
  \mathcal{L}_{\mathrm{int}} = \sum_\flabel \frac{C_\flabel}{2f_a} \left(
    \partial_\mu a
  \right) \overline{\psi}_\flabel \gamma^\mu \gamma_5 \psi_\flabel,
\end{align}
where $f_a$ is the axion decay constant and $C_\flabel$ are the model-dependent coefficients.
The index $\flabel$ labels the SM fermions, including electron $e$, neutron $n$, and proton $p$.\footnote{
    In the KSVZ axion model~\cite{PhysRevLett.43.103,SHIFMAN1980493} we have $|C_{e}|\ll 1$ while $C_{p} \sim C_{n} \sim O(1)$. In the DFSZ axion model~\cite{DINE1981199,Zhitnitsky:1980tq} or the flavorful axion model~\cite{Ema:2016ops,Calibbi:2016hwq}, we have $C_e \sim C_n \sim C_p \sim O(1)$.
}
In the non-relativistic limit, this interaction term describes the axion interaction with fermion spins $\vec{S}_\flabel$ given by
\begin{align}
  H_{\mathrm{int}} = \sum_\flabel \frac{g_{a\flabel\flabel}}{m_\flabel}
  \vec{\nabla} a \cdot \vec{S}_\flabel,
  \label{eq:a_f_interaction}
\end{align}
where the dimensionless coupling constants $g_{a\flabel\flabel} \equiv C_\flabel m_\flabel / f_a$ are used.
It is seen that the axion gradient $\vec\nabla a$ may be regarded as an effective magnetic field, and the interaction can be rewritten as
\begin{align}
  H_{\mathrm{int}} = \sum_\flabel \gamma_\flabel \vec{B}_\flabel(t) \cdot \vec{S}_\flabel,
\end{align}
with the gyromagnetic ratio $\gamma_\flabel$.
Substituting \cref{eq:coherent_oscillation} into the above expression, the fermion-dependent effective magnetic fields are calculated as
\begin{align}
  \vec{B}_\flabel(t) \simeq \frac{g_{a\flabel\flabel}}{m_\flabel \gamma_\flabel} \sqrt{2\rho_a} \vec{v}_a \cos (m_a t + \phi),
  \label{eq:b_f_vec}
\end{align}
where higher-order terms of $v_a \ll 1$ are neglected.
The amplitude of the effective magnetic field $B_\flabel^0$ is estimated as
\begin{align}
  B_\flabel^0 \sim \SI{4}{aT} \times \left(
    \frac{g_{a\flabel\flabel}}{10^{-10}}
  \right).
\end{align}
Note that, for typical axion models with $C_e \sim C_n \sim C_p$, this ``effective'' magnetic field for nucleons $B_{n,p}^0$ is larger by a factor $\sim m_{n,p}/m_e$ than that for the electron $B_e^0$.
Taking account of the gyromagnetic ratio for the nucleon/electron, it results in the same order of the interaction strengths with the electron and nucleons.
This is qualitatively different from the ordinary magnetic field, which acts on the electron spin much more strongly than on nucleons due to the difference of the gyromagnetic ratio.
In this sense, the use of nuclear spins in the NV center can offer complementary sensitivity in axion DM searches.

For isolated fermion spins, the corresponding $\vec{B}_\flabel(t)$ works just the same as the ordinary magnetic field aside from that its amplitude and oscillation phase vary with the time scale of $\tau_a$.
It is the effective magnetic field for electrons, $\vec{B}_e(t)$, that is sensed by the ordinary NV center magnetometry as proposed in \cite{Chigusa:2023hms}.
On the other hand, if we use nuclear spins for magnetometry, both $\vec{B}_n(t)$ and $\vec{B}_p(t)$ can be relevant as we will see shortly.

\subsection{Axion interaction with the $\Ntr$ spin}
\label{sec:axion_interaction}

The interaction between the $\Ntr$ spin and an ordinary magnetic field is characterized by its gyromagnetic ratio $\gamma_N$.
As a rare stable odd-odd nucleus, $\gamma_N$ has contributions from both the neutron and proton spins and the orbital angular momentum of the proton.
However, $\gamma_N$ does not reflect the axion interaction strength with the $\Ntr$ spin, which is determined by the axion interaction with neutron and proton spins, $g_{ann}$ and $g_{app}$.
To accurately describe the axion-$\Ntr$ interaction, we need to understand the composition of the $\Ntr$ spin $I=1$.
In this subsection, we discuss this issue under the assumption that the nuclear shell model well describes internal structure of the $\Ntr$ nucleus.

$\Ntr$ has seven neutrons and seven protons.
According to the nuclear shell model, both kinds of nucleons occupy the orbitals as $1s_{1/2}^2 1p_{3/2}^4 1p_{1/2}$, where both $1s_{1/2}$ and $1p_{3/2}$ orbitals form closed shells.
Thus, the nuclear spin $I=1$ comes from the synthesis of the total neutron spin $J_n=1/2$ of a neutron in the $1p_{1/2}$ orbital and $J_p=1/2$ of a proton in the proton counterpart of the orbital.
In the representation theory of $SU(2)$, this corresponds to the decomposition $\frac{1}{2} \otimes \frac{1}{2} = 1 \oplus 0$ with the first term in the right-hand side is selected, while each spin-$\frac{1}{2}$ representation on the left-hand side comes from the decomposition $\frac{1}{2}\otimes 1 = \frac{3}{2} \oplus \frac{1}{2}$.

Let us explicitly write down the $I=1$ states in terms of the eigenstates of the spin $\vec{S}_\flabel$ and the orbital angular momentum $\vec{L}_\flabel$ of nucleons $\flabel=n,p$ in the $1p$ orbitals.
Let $\ket{\uparrow}_\flabel$ and $\ket{\downarrow}_\flabel$ be the spin up and down states, and $\ket{m}_\flabel$ ($m=-,0,+$) be the eigenstates of $L_\flabel^z$ for each nucleon $f$.
Corresponding to this decomposition $\frac{1}{2}\otimes 1 = \frac{3}{2} \oplus \frac{1}{2}$, the $J_\flabel=\frac{1}{2}$ component (i.e., the $1p_{1/2}$ orbital for the nucleon $\flabel$) is given by
\begin{align}
  \begin{pmatrix}
    \ket{u_\flabel} \\
    \ket{d_\flabel}
  \end{pmatrix} \equiv \frac{1}{\sqrt{3}} \begin{pmatrix}
    \ket{\uparrow}_\flabel \ket{0}_\flabel - \sqrt{2} \ket{\downarrow}_\flabel \ket{+}_\flabel \\
    \sqrt{2} \ket{\uparrow}_\flabel \ket{-}_\flabel - \ket{\downarrow}_\flabel \ket{0}_\flabel
  \end{pmatrix}.
\end{align}
Using these expressions to evaluate the second decomposition $\frac{1}{2} \otimes \frac{1}{2} = 1 \oplus 0$, the nuclear spin $I=1$ component is expressed as
\begin{align}
  \vec{\psi}_I^\intercal \equiv
  \begin{pmatrix}
    \ket{u_p} \ket{u_n} \\
    \dfrac{1}{\sqrt{2}} \pqty{
      \ket{u_p} \ket{d_n} + \ket{d_p} \ket{u_n}
    } \\
    \ket{d_p} \ket{d_n}
  \end{pmatrix}.
  \label{eq:1ph_basis}
\end{align}

To go further, we calculate the matrix elements of the spin operators $\vec{S}_\flabel$ ($\flabel=n,p$) in the $36$-dimensional space corresponding to the all possible choices of $S_\flabel^z$ and $L_\flabel^z$ ($\flabel=n,p$).
In particular, since the nuclear spin $I=1$ states correspond to the three dimensional subspace spanned by three vectors in $\vec{\psi}_I$, matrix elements of the spin operators $\vec{S}_\flabel$ ($\flabel=n,p$) for these basis vectors represent how the axion interacts with the $\Ntr$ spin.
From a straightforward calculation, we obtain the following
\begin{align}
  \vec{\psi}_I^\dagger S^\alpha \vec{\psi}_I = -\frac{1}{6} I^\alpha, \label{eq:S_I_relation}
\end{align}
where $I^\alpha$ $(\alpha=x,y,z)$ are the spin-1 representations of the $SU(2)$ generators.
This result is consistent with the treatment in \cite{Ema:2024oce}.
Note that the spin operators also have non-zero matrix elements outside the three dimensional subspace, a part of which connects different spin states.
These interactions can in principle invoke the transition from the ground state with $I=1$ to, e.g.\ an excited state with $I=0$.
However, since the relevant energy scale of $O(1$--$10)\,\si{MeV}$~\cite{PhysRev.130.1530} is far beyond the current setup, we can safely neglect these terms and focus on the terms in \cref{eq:S_I_relation} that preserve the nuclear spin structure.

Since the right-hand side of \cref{eq:S_I_relation} is proportional to the nuclear spin operators $I^\alpha$, $\vec{S}_\flabel$ ($\flabel=n,p$) effectively acts as the $I=1$ spin operators with a non-trivial coefficient $-1/6$.
In \cref{sec:synthesis}, we provide proof that the same interpretation is possible whenever the spin $S=1/2$ and a general orbital angular momentum $L=\ell$ are considered, and derive a systematic way to calculate the coefficient.
By substituting \cref{eq:S_I_relation} in \cref{eq:a_f_interaction}, we obtain an effective axion-$\Ntr$ interaction Hamiltonian
\begin{align}
  H_{\mathrm{int}} = \gamma_N \vec{B}_N(t) \cdot \vec{I},
\end{align}
with the effective magnetic field defined as
\begin{align}
  \vec{B}_N(t) &\equiv B_N \hat{v}_a \cos(m_a t + \phi), \label{eq:b_n_vec} \\
  \gamma_N B_N &\simeq - \frac{1}{6} \left(
    \frac{g_{app}}{m_p} + \frac{g_{ann}}{m_n}
  \right) \sqrt{2\rho_a} v_a,
  \label{eq:gamma_B_N}
\end{align}
with $\hat{v}_a \equiv \vec{v}_a / v_a$.
For convenience, we define
\begin{align}
  \tilde{f}_a \equiv \left|
    \frac{g_{app}}{2m_p} + \frac{g_{ann}}{2m_n}
  \right|^{-1},
  \label{eq:f_a_tilde}
\end{align}
with which $B_N \propto \tilde{f}_a^{-1}$.
Since we can rewrite it as $\tilde{f}_a = 2f_a / (C_p + C_n)$, $\tilde{f}_a$ is of the same order as $f_a$ if coefficients $C_p$ and $C_n$ are of $O(1)$.
A fascinating consequence of the $\Ntr$ spin as an odd-odd nucleus is that the axion coupling is proportional to the combination \cref{eq:f_a_tilde}, and that the coupling strength is sensitive to the relative sign of $g_{app}$ and 
$g_{ann}$.

\section{Sensitivity estimation}
\label{sec:sensitivity}

When measurements are repeated $\Nobs$ times, we obtain time-sequence data labeled by $j=1,\dots,\Nobs$, representing the measurement starting at time $t_j$.
For simplicity, we assume $t_j=(j-1) \tau$ with $\tau = T_{2n}^*/2$ ($T_{2n}/2$) for the dc (ac) effective magnetometry approach,\footnote{
In this context, we neglect the measurement overhead, including the state preparation and fluorescence measurement.
This is a reasonable approximation given that $\tau \sim O(1)\,\si{ms}$ while the overhead time is typically of $O(10$--$100)\,\si{\mu s}$ \cite{Barry:2023hon}.
} though it is not necessary for the following discussion that the measurements are repeated with equal time intervals.
As a result, $\tobs \equiv \Nobs \tau$ denotes the total observation time.
Let $\rho_j$ be the density matrix representing the quantum state of the NV center ensemble before the $j$-th fluorescence measurement.
Since our observable is defined as an operator
\begin{align}
  M_j^z \equiv \frac{1}{2N} \sum_{\ell=1}^N \sigma_{j\ell}^z,
\end{align}
where $\sigma_{j\ell}^z$ acts on the qubit state of the nuclear spin in the $\ell$-th NV center at the $j$-th measurement, the data obtained at $t_j$ can be calculated as $\Braket{M_j^z}_{\rho_j} \equiv \MyTr{\rho_j M_j^z}$.
It should be noted that $\Braket{M_j^z}_{\rho_j}$ represents the $N$-qubit average of the signal at time $t_j$, which asymptotes to the signal strength $F$ with an appropriate choice of the phase factor in the limit of $N\to \infty$.

The expression above is useful for calculating the ensemble average over distributions of the axion parameters.
It should be noted that $\rho_j$ is equivalent to \cref{eq:qs_dc} and to \cref{eq:qs_ac} with the replacement $\phi \to m_a t_j + \phi$ for the Ramsey and the Hahn echo sequences, respectively. 
Therefore, it depends on the axion velocity $\vec{v}_a$ and the phase factor $\phi$ through the expression of the effective magnetic field \cref{eq:gamma_B_N}.
If we neglect the Earth's relative motion against the Galactic center, $\phi$ is uniformly distributed in the range $[0, 2\pi)$, while the axion velocity has a random direction with typically a value of $v_a \sim 10^{-3}$.
Using these distributions, for example, the ensemble average of the observation result $\mathcal{M}_j$ is calculated as
\begin{align}
  \Braket{\mathcal{M}_j} \equiv \left. \frac{1}{2\pi} \int d\phi\,
  \frac{1}{4\pi} \int d\hat{v}_a\,
  \MyTr{
    \rho_j M_j^z
  } \right|_{v_a = 10^{-3}},
  \label{eq:ensemble_av}
\end{align}
where we do not take into account the distribution of $v_a$, which highly depends on the model of the DM profile in our galaxy and only results in an $O(1)$ modification.
It should be noted that, here and hereafter, we neglect the subscript of the ensemble-averaged quantities for notational simplicity.
As anticipated, the randomness of the signal direction and phase causes cancellation of the averaged signal, $\Braket{M_j} = 0$.

To derive meaningful insights from the data, we use two-point functions of the time-sequence data defined as~\cite{Dror:2022xpi}
\begin{align}
  C_{jj'} \equiv
  \begin{cases}
    \MyTr{(\rho_j \otimes \rho_{j'})(M_j^z \otimes M_{j'}^z)}, & (j\neq j') \\
    \MyTr{\rho_j M_j^z M_j^z}. & (j=j')
  \end{cases}
\end{align}
Since the coherence of the axion signal is maintained only for the duration $\tau_a$, $C_{jj'}$ behaves differently for $|t_j - t_{j'}| < \tau_a$ and $|t_j - t_{j'}| > \tau_a$.
A combined expression can be given as 
\begin{widetext}
\begin{align}
  C_{jj'} = &\left. \frac{1}{2\pi} \int d\phi\, \frac{1}{4\pi} \int d\hat{v}_a\,
  \frac{1}{2\pi} \int d\phi'\, \frac{1}{4\pi} \int d\hat{v}_a'\,
  \MyTr{
    \rho_{jj'} M_{jj'}^z
  } \right|_{v_a = 10^{-3}} \notag \\
  &\times \left[
    \Theta(|t_j - t_{j'}| - \tau_a) + 8\pi^2 \delta(\phi-\phi') \delta(\hat{v}_a - \hat{v}_a') \Theta(\tau_a - |t_j - t_{j'}|)
  \right],
  \label{eq:Ctt}
\end{align}
\end{widetext}
where $\rho_{jj'} \equiv \rho _j \otimes \rho _{j'}$ and $M_{jj'}^z \equiv M_j^z \otimes M_{j'}^z$ for $j\neq j'$ and $\rho_{jj} \equiv \rho _j$ and $M_{jj}^z \equiv M_j^z M_{j}^z$.
Also, $\Theta$ is the Heaviside step function.
The integral variables $(\phi, \hat{v}_a)$ and $(\phi', \hat{v}_a')$ correspond to the axion parameters at time $t_j$ and $t_{j'}$, respectively, and the delta functions in the second line denote the coherence of the signal for $|t_j - t_{j'}| < \tau_a$.
In addition, we introduce the power spectral density (PSD), which is defined as the ensemble-averaged expectation value $\PSD{k} \equiv \Braket{\mathcal{O}_k}$ of the operator
\begin{align}
  \mathcal{O}_k \equiv \frac{\tau^2}{\tobs} \sum_{j,j'} e^{2\pi i k (j-j')/\Nobs} M_j^z M_{j'}^z,
  \label{eq:Ok_def}
\end{align}
with $k=0,\dots,\Nobs-1$.
Each $\PSD{k}$ can be calculated through the Fourier transformation of the two-point functions as
\begin{align}
  \PSD{k} = \frac{\tau^2}{\tobs} \sum_{j,j'} e^{2\pi i k (j-j')/\Nobs} C_{jj'}.
  \label{eq:PSD_def}
\end{align}

A detailed calculation of the PSD and the relevant noise contributions is given in \cref{sec:PSD}.
From \cref{eq:PSD_exp}, the signal PSD can be defined as $\SPSD{k} \equiv \PSD{k} - \tau/(4N)$, where the constant shift ensures that $\SPSD{k}$ is proportional to the axion-induced magnetic field $B_N$, and is given by
\begin{align}
  \SPSD{k} \simeq \dfrac{2\mathcal{A}}{\tobs \Delta\omega_k^2} \sin^2 \dfrac{\tobs \Delta\omega_k}{2},
\end{align}
for $\tobs < \tau_a$, while
\begin{align}
  \SPSD{k} \simeq \dfrac{2\mathcal{A}}{\tobs \Delta\omega_k^2} \sin^2 \dfrac{\tau_a \Delta\omega_k}{2} + \dfrac{\tobs-\tau_a}{\tobs \Delta\omega_k} \mathcal{A} \sin\left[
    \tau_a \Delta\omega_k
  \right],
\end{align}
for $\tobs > \tau_a$, where $\Delta\omega_k \equiv \omega_k - m_a$ with $\omega_k \equiv 2\pi k/\tobs$ and $\mathcal{A} \propto B_N^2$ is the protocol-dependent coefficient defined in \cref{eq:A}.
Due to the quantum noise, the measurement result of the PSD fluctuates even without the axion DM.
The standard deviation of the PSD distribution is calculated as
\begin{align}
  \BPSD{k} \equiv \left.
    \sqrt{\Braket{\mathcal{O}_k^2} - \Braket{\mathcal{O}_k}^2}
  \right|_{B_N = 0}.
  \label{eq:dPSD}
\end{align}
As shown in \cref{eq:Bk_exp}, we obtain $\BPSD{0} \simeq \tau/(2\sqrt{2}N)$ and $\BPSD{k\neq 0} \simeq \tau/(4N)$
for our setup.

Focusing on a single bin $k$, the signal estimation uncertainty can be evaluated through the well-known formula
\begin{align}
  \delta B_N^2 = \left. \sqrt{\Braket{\mathcal{O}_k^2} - \Braket{\mathcal{O}_k}^2} \left(
    \frac{d\Braket{\mathcal{O}_k}}{d B_N^2}
  \right)^{-1} \right|_{B_N^2 = 0},
  \label{eq:single-bin}
\end{align}
which determines the estimation error of $B_N^2$ around the specific choice of $B_N^2 = 0$, i.e.\ the model without the axion DM.
We select $B_N^2$ as a parameter to be estimated since $\Braket{O_k}$ does not have a linear term in $B_N$ as shown in \cref{eq:PSD_exp}.
By deforming the above expression with the relationship $\delta B_N^2 = 2B_N \delta B_N$, we can obtain the $X\sigma$-level detection limit to the axion-induced magnetic field $B_N$ to be $X\delta B_N$ (where $X$ depends on the required confidence level).
To gather information from all bins and obtain the best achievable detection limit with these observables, we follow~\cite{Dror:2022xpi}.
Based on the Asimov dataset~\cite{Cowan:2010js} rather than the Monte Carlo simulation results, we compute the test statistic
\begin{align}
  q = 2\sum_{k=0}^{\Nobs-1} \left[
    \left(
      1 - \frac{\BPSD{k}}{\SPSD{k} + \BPSD{k}}
    \right)- \ln \left(
      1 + \frac{\SPSD{k}}{\BPSD{k}}
    \right)
  \right],
\end{align}
with which the $95\,\%$ exclusion limit, which we adopt as the definition of the detection limit of our approach, is determined by the criteria $q=-2.71$.

It is beneficial to consider two extreme setups and evaluate the scaling of the detection limit as a function of $\tobs$ and $N$.
For this purpose, we first observe that the signal PSD $\SPSD{k}$ has a resonant structure peaked at $\Delta\omega_k=0$ or $\omega_k = m_a$.
The peak height is evaluated as
\begin{align}
  \SPSD{k} = \begin{cases}
    \mathcal{A} \dfrac{\tobs}{2}, & (\tobs < \tau_a) \\[10pt]
    \mathcal{A} \tau_a \left(
      1 - \dfrac{\tau_a}{2\tobs}
    \right), & (\tobs > \tau_a)
  \end{cases}
\end{align}
where the resonance condition for these setups can be described as $\tobs\Delta\omega_k \ll 1$ and $\tau_a\Delta\omega_k \ll 1$, respectively.
When $\tobs \ll \tau_a$, the signal peak height grows linearly with $\tobs$, while only a single bin enjoys resonance since $\tobs(\omega_{k+1}-\omega_k) = 2\pi$.
Then, we can approximate the test statistic as
\begin{align}
  q \simeq 2\left(
    1 - \frac{\BPSD{k_0}}{\SPSD{k_0} + \BPSD{k_0}}
  \right) - \ln \left(
    1 + \frac{\SPSD{k_0}}{\BPSD{k_0}}
  \right),
\end{align}
with $k_0$ being the label of the resonance bin.
Since the above expression only depends on the ratio $\SPSD{k_0}/\BPSD{k_0}$, the detection limit is solely determined by solving $q=-2.71$ for this ratio, resulting in $\SPSD{k_0}/\BPSD{k_0} \simeq 8.48$.
Since $\SPSD{k_0}/\BPSD{k_0}$ is proportional to $N \tobs B_N^2$, the detection limit to $B_N$ grows as $N^{1/2} \tobs[1/2]$ as expected for a coherently oscillating signal.
On the other hand, when $\tobs \gg \tau_a$, the peak height is saturated to $\sim \mathcal{A} \tau_a$, but the number of bins involved in the peak grows as $\tobs/\tau_a$.
In this limit, our setup can be sensitive to small signals with $\SPSD{k} \ll \BPSD{k}$, where we can expand the expression of the test statistic as
\begin{align}
  q \simeq -\sum_k \frac{\SPSD[2]{k}}{\BPSD[2]{k}}.
\end{align}
Since the number of terms with dominant contributions grow as $\tobs$ and the fraction in the summation is proportional to $N^2 B_N^4$, we obtain the detection limit scaling $\propto N^{1/2} \tobs[1/4]$.
Again, this scaling behavior is common for the signal with randomized direction and phase.

To summarize, the sensitivity to the axion coupling is roughly estimated from 
    \begin{align}
  \sqrt\mathcal {A} \sim \begin{cases}
    \sigma_R \left(\dfrac{\tau}{N \tobs}\right)^{1/2}, & (\tobs < \tau_a) \\
    \sigma_R \left(\dfrac{\tau}{N \tau_a}\right)^{1/2}\left(\dfrac{\tau_a}{\tobs}\right)^{1/4}, & (\tobs > \tau_a)
  \end{cases}
  \label{eq:sensitivity}
\end{align}
where $\mathcal A$ is defined in \cref{eq:A}.
$\sigma_R$ parametrizes the size of the shot noise as detailed in \cref{sec:shot_noise}; $ \sigma_R = 1$ corresponds to the ideal case when the sensitivity is limited by the projection noise, while $\sigma_R \simeq 19$ has been already achieved~\cite{Barry:2023hon}, and is expected to reduce further.

The analysis explained so far uses the full data set with $j=1,\dots,\Nobs$ and their Fourier transformation to look for a signal.
However, this should be interpreted as a way to estimate the best achievable detection limit curves.
In realistic experimental setups, on the other hand, there are several challenges to performing such an analysis including memory constraints and limitations on computational power.
Given these constraints and limitations, an alternative analysis procedure is the one based on the standard deviation~\cite{SDQS}.
We can show that, by setting an appropriate data collection time duration, the sensitivities of this procedure to the target frequencies have the same scaling behavior with $\tobs$ and $N$ as shown above.

\section{Results}
\label{sec:results}

\begin{figure}[t]
  \centering
  \includegraphics*[width=\hsize, page=1]{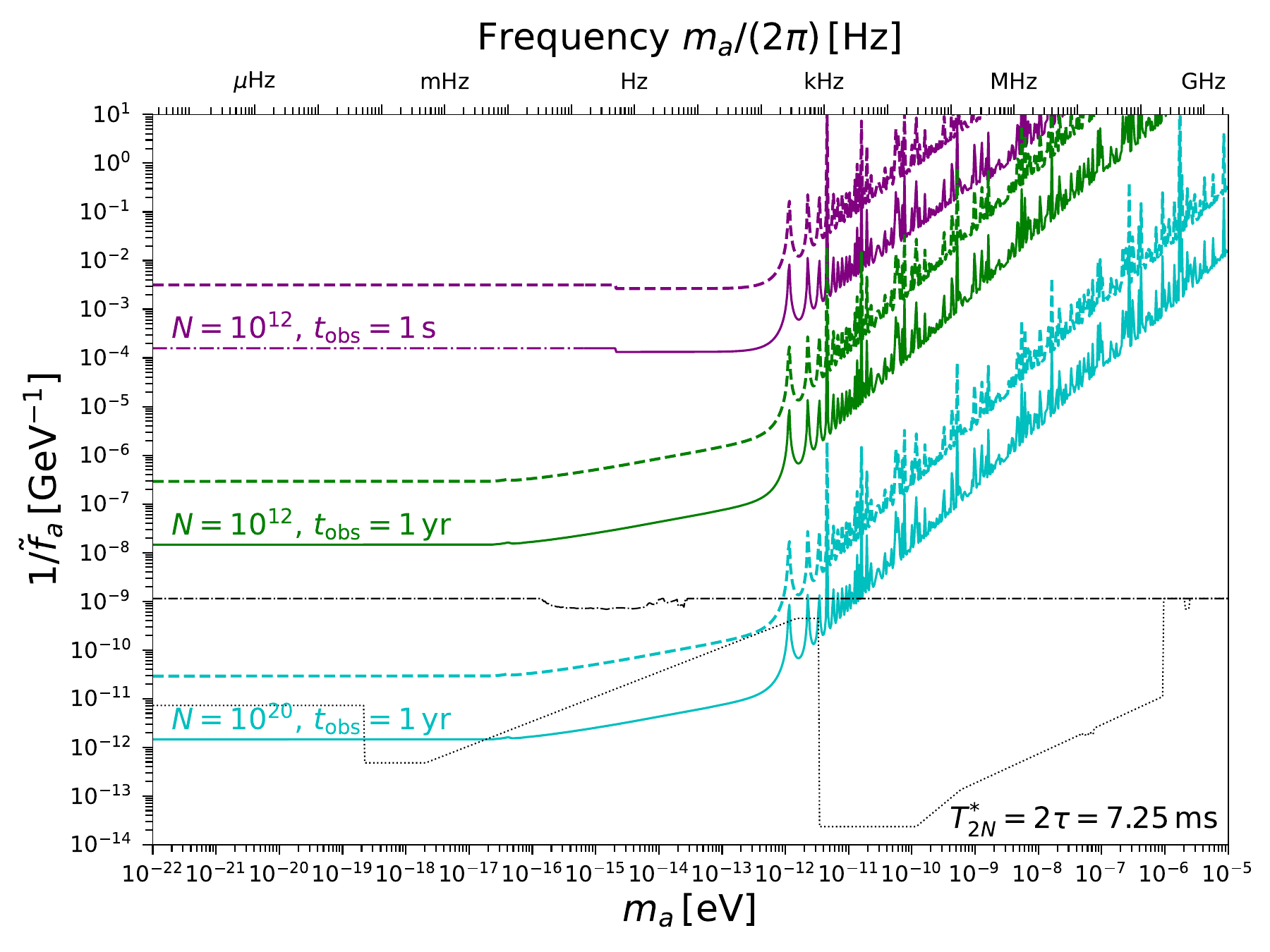}
  \caption{
    The calculated $95\,\%$ exclusion limits on $\tilde{f}_a$ as a function of $m_a$ for $T_2^{*} = 2\tau = \SI{7.25}{ms}$.
    The total detector volumes of $(N, \tobs)=(10^{12}, \SI{1}{s})$ (magenta), $(10^{12}, \SI{1}{yr})$ (green), and $(10^{20}, \SI{1}{yr})$ (cyan) are assumed.
    The colored solid (dashed) lines represent the projection noise-limited (the shot noise-limited) sensitivities with $\sigma_R = 1$ ($\sigma_R = 20$).
    The black dash-dotted line represents the combination of the current best constraints on $|g_{app}|$ and $|g_{ann}|$, including constraints on $|g_{ann}|$ from neutron star cooling~\cite{Buschmann:2021juv}, $\ce{K}$--$\ce{^3He}$ comagnetometer~\cite{vasilakis2009limits}, and ChangE~\cite{Wei:2023rzs} and ChangE NMR~\cite{Xu:2023vfn} experiments, and a constraint on $|g_{app}|$ from SN1987A~\cite{Lella:2023bfb}.
    The black dotted line represents the prospect of constraints, including constraints on $|g_{ann}|$ from future comagnetometers~\cite{Bloch:2019lcy,Wu:2019exd}, the electrostatic storage ring~\cite{Brandenstein:2022eif}, the CASPEr-gradient experiment~\cite{JacksonKimball:2017elr}, and the homogeneous precession domain of the superfluid $\ce{^3He}$~\cite{Gao:2022nuq}, and constraints on $|g_{app}|$ from the proton storage ring~\cite{Graham:2020kai}, the CASPEr-gradient experiment~\cite{JacksonKimball:2017elr}, and the nuclear magnon in $\ce{MnCO3}$~\cite{Chigusa:2023hmz}.
    The limit data is adopted from \cite{OHare:2020}.
  }
  \label{fig:dc}
\end{figure}

\begin{figure}[t]
  \centering
  \includegraphics*[width=\hsize]{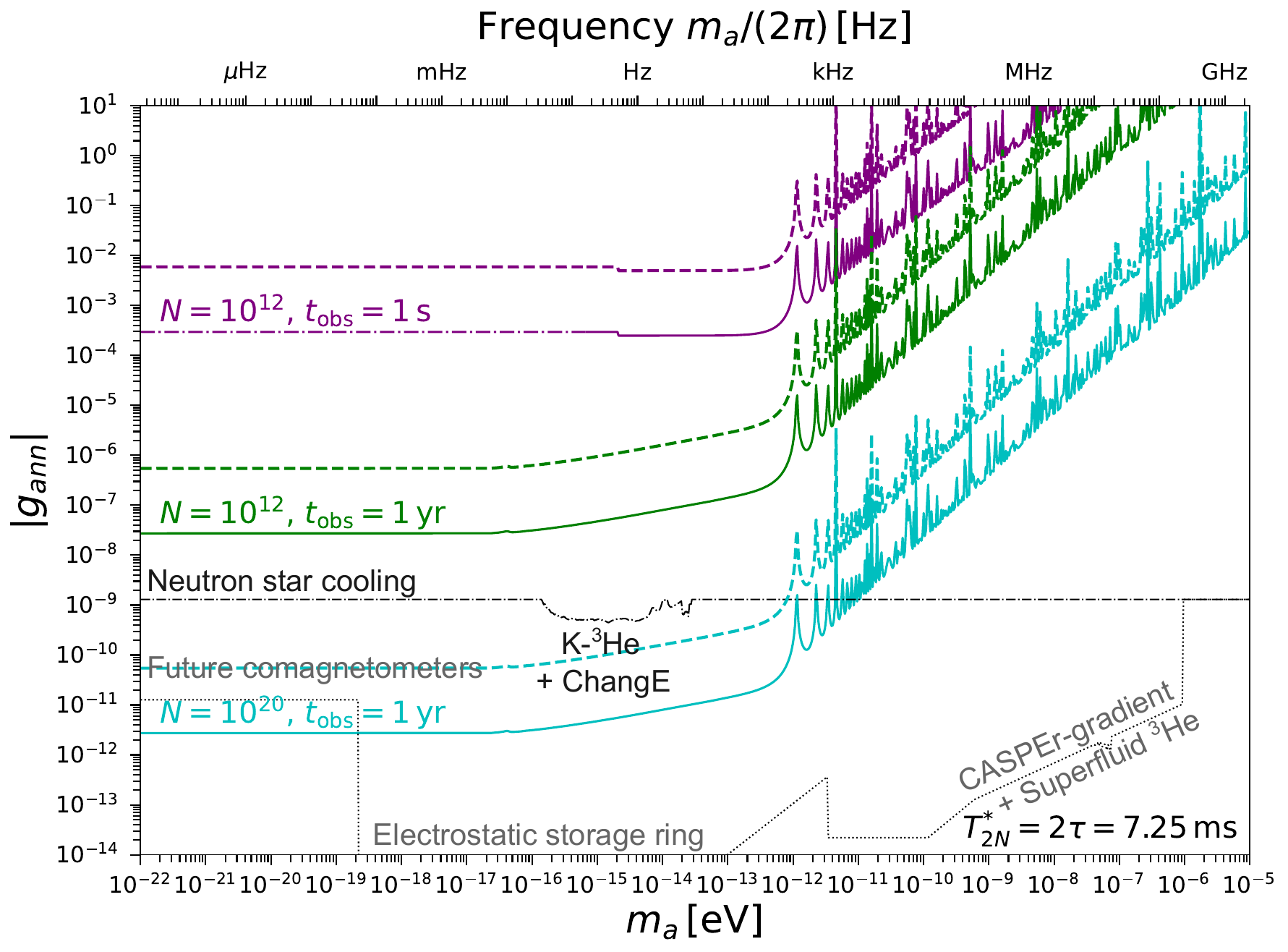}
  \includegraphics*[width=\hsize]{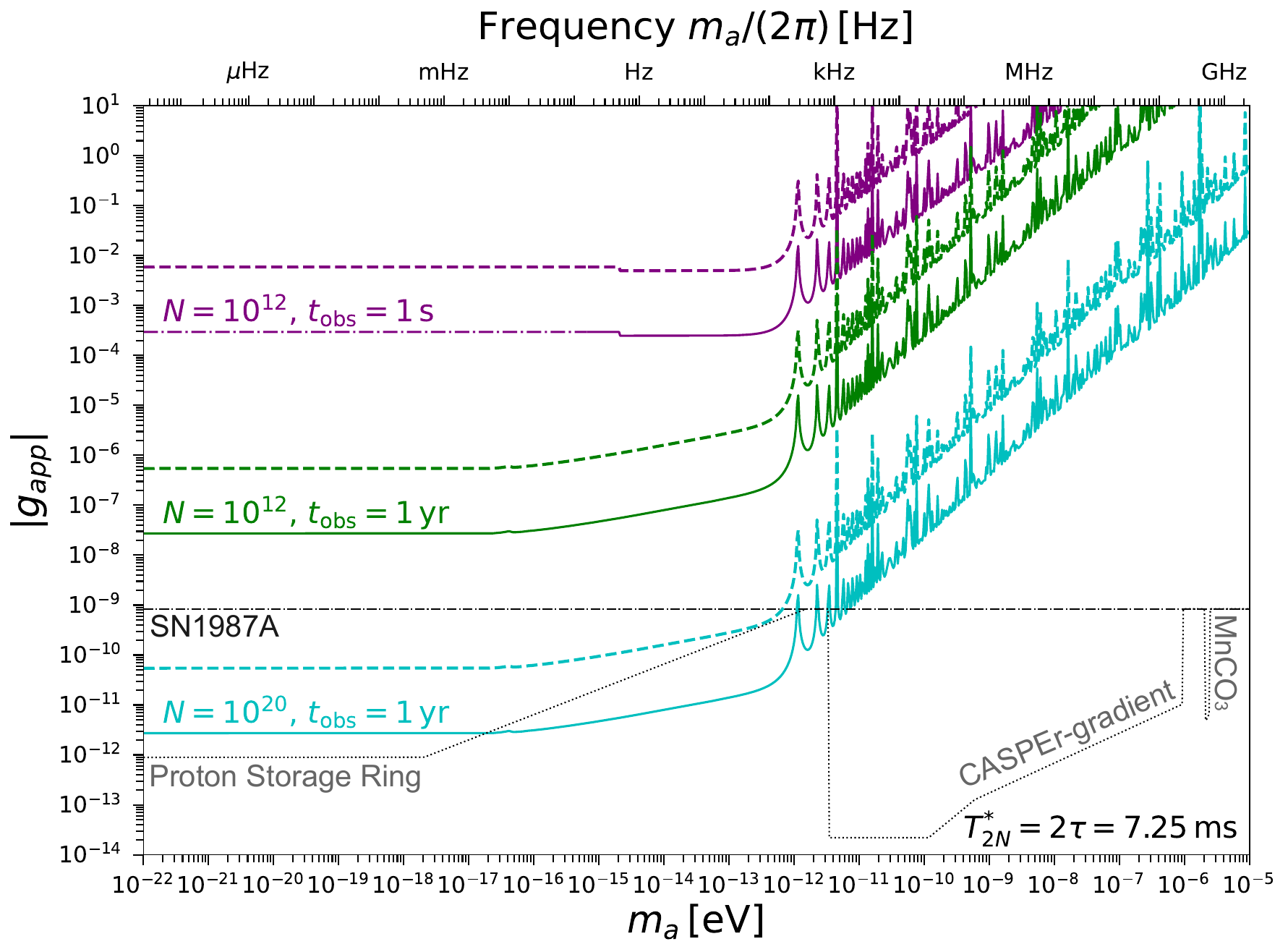}  
  \caption{
    The top (bottom) panel shows the calculated limits on $|g_{ann}|$ ($|g_{app}|$) from the Ramsey setup under an artificial assumption that $|g_{app}| \ll |g_{ann}|$ ($|g_{ann}| \ll |g_{app}|$).
    We assume the dephasing time scale of $T_2^{*} = \SI{7.25}{ms}$.
    The color conventions and the meaning of the black lines for the existing constraints and prospects are the same as in \cref{fig:dc}.
  }
  \label{fig:dc_individual}
\end{figure}

The calculated $95\,\%$ exclusion limits on $\tilde{f}_a$ defined in \cref{eq:f_a_tilde} from the Ramsey sequence are shown in \cref{fig:dc} with the assumed relaxation time and the free precession time $T_{2N}^{*} = \SI{7.25}{ms}$, choosing $\tau = T_{2N}^{*}/2$.\footnote{
Precisely speaking, there are periodic $O(1)$ fluctuations of the sensitivity due to the discrete binning of the frequency with the bin width $2\pi/\tobs$.
In \cref{fig:dc}, we smooth out these fluctuations to focus on the larger-scale frequency dependence of the sensitivity.
Furthermore, the small step of the magenta line at $\sim \SI{1}{Hz}$ is due to the difference between \cref{eq:B0_exp,eq:Bk_exp}.
}
Three colored lines correspond to the most conservative setup with an already-achieved number of NV centers $N=10^{12}$~\cite{Barry:2023hon} and $\tobs = \SI{1}{s}$ (magenta), the same $N=10^{12}$ but with $\tobs = \SI{1}{yr}$ (green), and a rather optimistic choice of $N=10^{20}$ with $\tobs = \SI{1}{yr}$ (cyan).
The solid and dash-dotted lines represent the projection noise-limited sensitivities ($\sigma_R = 1$), while the dashed lines the shot noise-limited sensitivities with the choice of $\sigma_R = 20$.
When using the exact sample from the current state of NV sensors~\cite{Barry:2023hon}, which has approximately \SI{1}{mm} sides and a \SI{70}{\mu m} thickness, about $3\times10^8$ diamond samples are required to reach $N=10^{20}$ NV centers. This is a rather large number, however, there are many ways to decrease it. For example, improving the yield of NV center creation, here $0.68\%$, to the current state-of-the-art, which is $25.8\%$~\cite{Balasubramanian:2022}, reduces the required samples to about $8\times10^6$. Further decrease is possible by increasing sample thickness, or side dimensions if practically allowed. Improving the sensitivity, for example by lowering the temperature, using double quantum sequences, or increasing collection efficiency, can push this number down significantly as well, since the sensitivity is inversely proportional to the square root of the number of NV centers and thus samples (so a $10$ times better sensitivity means $100$ times less NVs are required). Hence, such enhancements bring a potential experiment into palatable proportions of other particle physics experiments.
Also shown by the black lines are the combined constraints on $\tilde{f}_a$ from the existing experimental results (dash-dotted) and the prospects (dotted).
See~\cite{OHare:2020} for details.

Since the Ramsey sequence delivers full performance when the frequency $f$ satisfies $f\lesssim 1/\tau$, the choice of $T_{2N}^{*} = 2\tau = \SI{7.25}{ms}$ leads to a frequency coverage $m_a/2\pi \lesssim \SI{200}{Hz}$ ($m_a \lesssim 8\times 10^{-13}\,\mathrm{eV}$), outside of which the sensitivity is rapidly lost.
Another remarkable feature of our sensitivities is the kinks of the green and cyan lines at $m_a \sim \SI{2e-17}{eV}$ with the corresponding axion coherence time $\tau_a \sim \SI{1}{yr}$.
Both lines below this point (and also the magenta line) correspond to $\tobs < \tau_a$, so the axion signal maintains coherence during the observation.
Thus, the Ramsey sequence has frequency-independent sensitivities for this mass range.
For higher masses, on the other hand, we need to account for a slower sensitivity improvement $\propto (\tau_a \tobs)^{1/4}$ shown in \cref{eq:sensitivity}, so the detection limit plots have slopes.
Finally, compared with solid lines that show detection limit prospects, the dash-dotted part of the magenta lines, which corresponds to the mass range $2\pi / m_a \ll \SI{1}{s}$, needs special care.
In this mass range, the signal strengths for all repeated measurements are proportional to $\cos\phi$ with a randomly chosen phase factor $\phi$; thus, it is always possible that no signal is observed irrespective of the size of $\tilde{f}_a$.
The magenta dash-dotted lines should then be interpreted as a $1\sigma$ lower bound on $\tilde{f}_a$ when no signal is observed.

Comparison between our results and the existing constraints or prospects shows that our approach is promising for a broad mass range with $m_a/2\pi \lesssim \SI{200}{Hz}$.
It should be noted, however, that the exclusion limits in \cref{fig:dc} need to be carefully interpreted since both previous results (black dashed lines) and other unrealized proposals (black dotted lines) have a dominant constraint on either $|g_{ann}|$ or $|g_{app}|$, contrary to our approach where $\tilde{f}_a$, a linear combination shown in \cref{eq:f_a_tilde}, is directly constrained.
Due to the expression \cref{eq:f_a_tilde}, both constraints on $|g_{ann}|$ and $|g_{app}|$ in principle affect the black lines.
However, practically, either the $|g_{ann}|$ or $|g_{app}|$ that is less constrained at a chosen $m_a$ determines how strongly $\tilde{f}_a$ is constrained at that value of $m_a$.
To disentangle the mixed effect of $|g_{ann}|$ and $|g_{app}|$ constraints, we also demonstrate the possible detection limits of our setup with $T_{2N}^{*} = \SI{7.25}{m s}$ on an individual $|g_{ann}|$ ($|g_{app}|$) coupling in the top (bottom) panel of \cref{fig:dc_individual} under an artificial assumption that the corresponding coupling is much larger than the other one.
\cref{fig:dc_individual} is useful for comparison; in particular, the optimistic setup (cyan) shows remarkable sensitivities to $|g_{ann}|$ for $m_a \lesssim \SI{2e-19}{eV}$ comparable to the future comagnetometer prospect~\cite{Bloch:2019lcy} and to $|g_{app}|$ for $\SI{e-15}{eV} \lesssim m_a \lesssim \SI{e-12}{eV}$ corresponding to a gap between the proton storage ring~\cite{Graham:2020kai} and the CASPEr-gradient prospects~\cite{JacksonKimball:2017elr}.
However, it should be remembered that \cref{fig:dc} is a more fundamental result of our approach obtained without any artificial assumptions on physics parameters.

\begin{figure}[t]
  \centering
  \includegraphics*[width=\hsize, page=1]{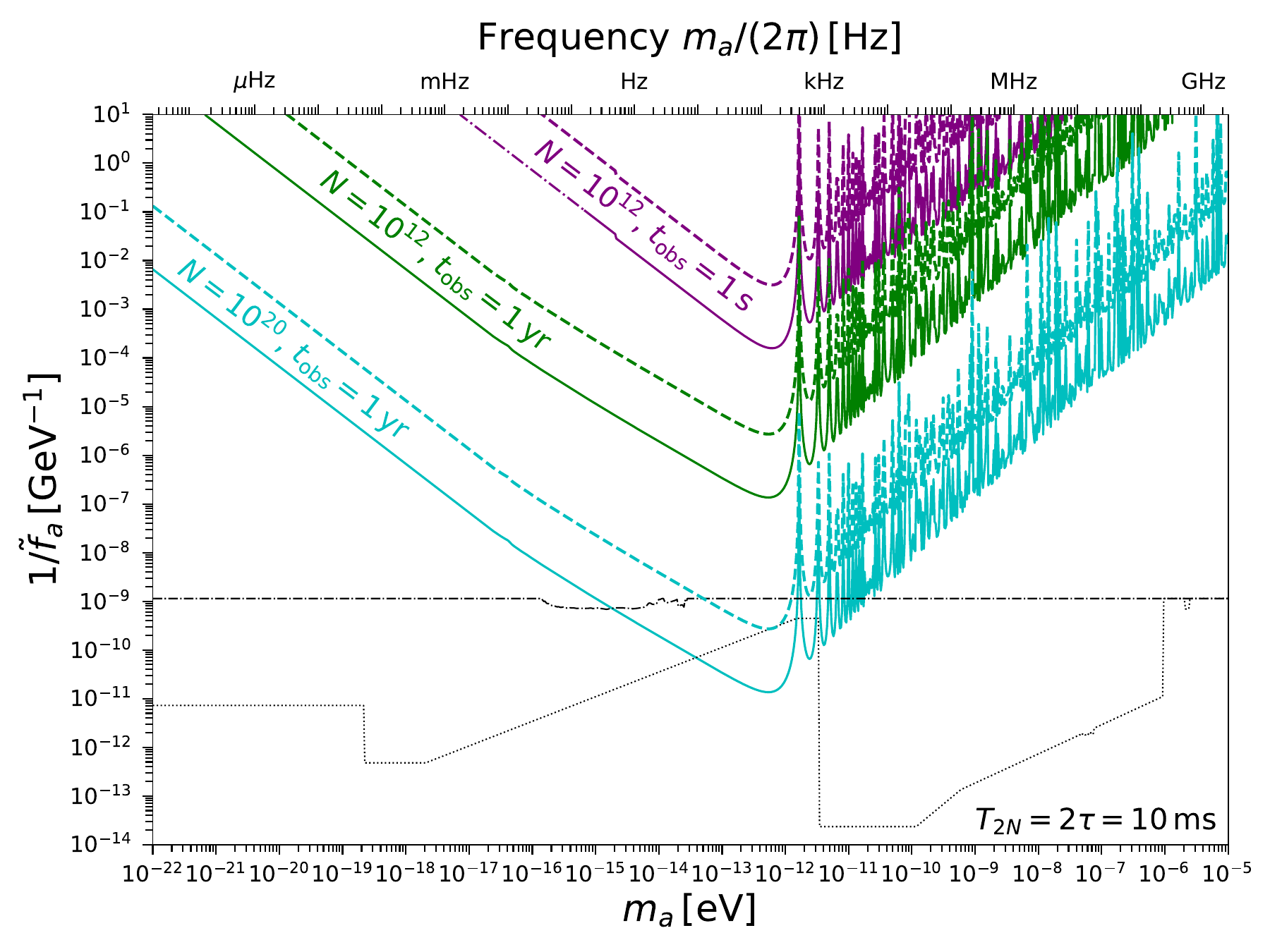}
  \includegraphics*[width=\hsize, page=2]{ac.pdf}
  \caption{
    Same as~\cref{fig:dc} but with the Hahn-echo sequence.
    The decoherence times of $T_2=\SI{10}{ms}$ (top) and $\SI{1}{s}$ (bottom) are assumed.
  }
  \label{fig:ac}
\end{figure}

In \cref{fig:ac}, we show the calculated $95\,\%$ exclusion limits from the Hahn-echo sequence at room temperature with $T_{2N} = \SI{10}{ms}$ (top) and a cryogenic environment with $T_{2N} = \SI{1}{s}$ (bottom).
As is clearly shown in the plots, this approach is a narrow-band search targeted at the frequencies $1/\tau \sim \SI{1}{Hz}$--$\SI{100}{Hz}$ depending on the choice.
Although the frequency coverage is limited in this approach, the sensitivity around the target frequency is much better than the Ramsey setup under the cryogenic environment when $T_{2N} \gg T_{2N}^{*}$.
Note that these exclusion limits can also be reinterpreted as limits on $|g_{ann}|$ and $|g_{app}|$ under certain assumptions similar to \cref{fig:dc_individual}.

\begin{figure}[t]
  \centering
  \includegraphics*[width=\hsize]{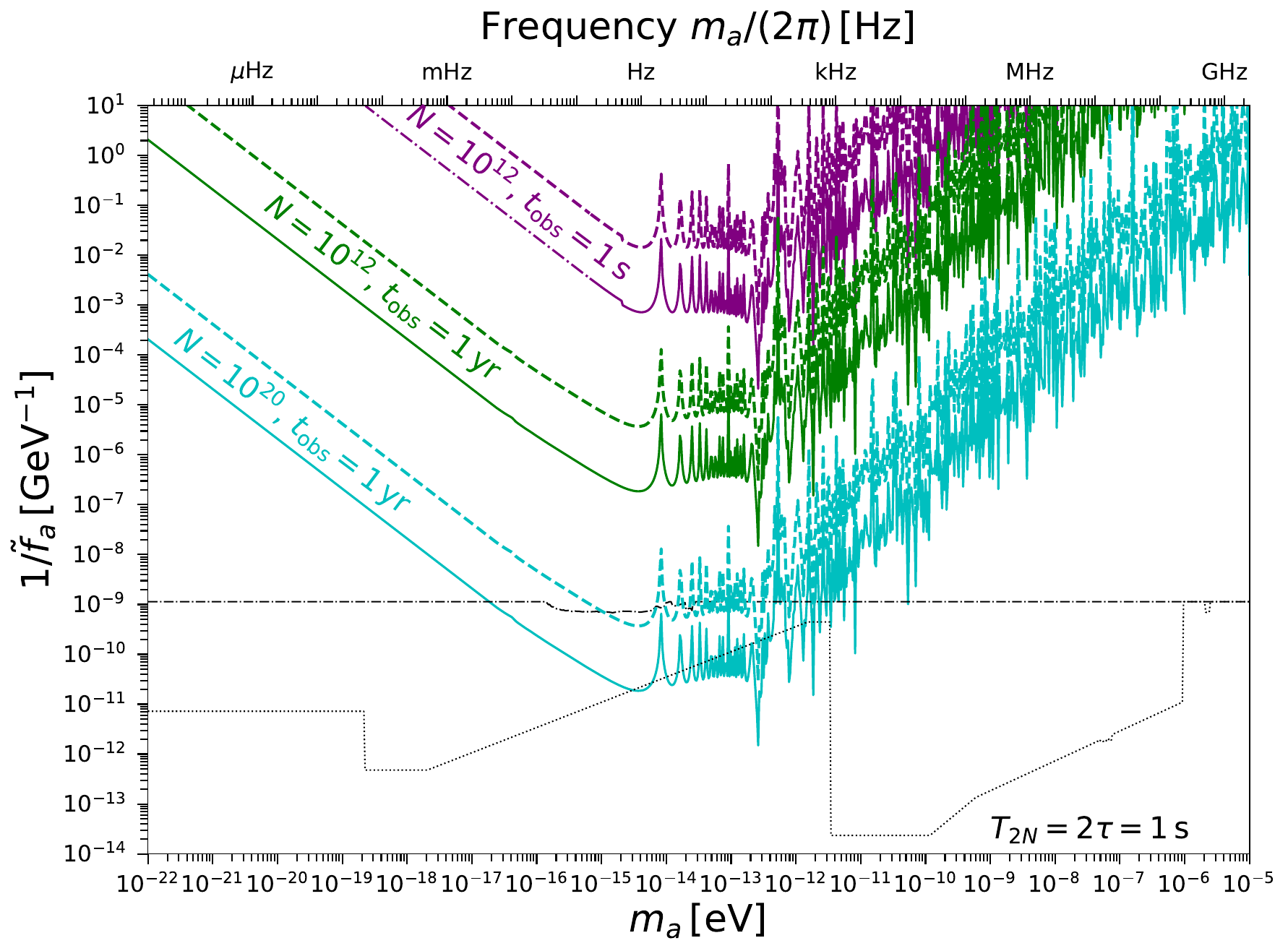}
  \caption{
    Same as~\cref{fig:dc} but with the DD sequence.
    The decoherence time of $T_2 = \SI{1}{s}$ and the number of $\pi$-pulses $N_\pi = 63$ are assumed.
  }
  \label{fig:dd}
\end{figure}

In \cref{fig:dd}, we show the calculated $95\,\%$ exclusion limits from the DD sequence with $T_2 = \SI{1}{s}$ and $N_\pi = 63$.
Despite the improved sensitivity at the peak due to the prolonged $T_2$, the peak width becomes narrower for a larger $N_\pi$, which makes this sequence generally not suitable for dark matter searches with unknown signal frequency. 

\section{Discussion and conclusion}
\label{sec:conclusion}

We proposed a novel method to use the $\Ntr$ spin of NV centers in diamond for axion dark matter searches.
Our nuclear spin magnetometry metrology approach is based on new types of protocols from \cref{fig:protocol_DC,fig:protocol_AC} aimed at dark matter searches, and provides potential constraints on the axion-nucleus couplings $g_{ann}$ and $g_{app}$, which are completely independent of those on $g_{aee}$ obtained with conventional magnetometry protocols in~\cite{Chigusa:2023hms}.
This opens up a new direction for quantum sensing techniques based on NV centers and motivates further investigation into the properties of the $\Ntr$ spin, including the relaxation time scales $T_{2N}^*$ and $T_{2N}$, for an ensemble of NV centers under various conditions.

One of the benefits of our approach compared with other proposed ideas to constrain $g_{ann}$ and $g_{app}$ is its broad frequency coverage arising from the wide dynamic range of NV center magnetometry.
We found several frequency windows in which our approach has a relatively high potential within the overall target frequency range $\lesssim \SI{100}{Hz}$.
Another remarkable feature is the natural sensitivity to a roughly $1:1$ linear combination of two coupling constants $g_{ann}$ and $g_{app}$ shown in \cref{eq:f_a_tilde}.
Accordingly, our approach is sensitive not only to an individual $g_{ann}$ or $g_{app}$ coupling under existence of a large hierarchy between them but also to a relative phase between them when they have comparable sizes, which enables us to explore the axion model after its discovery.

\begin{figure}[t]
  \centering
  \includegraphics[width=\hsize]{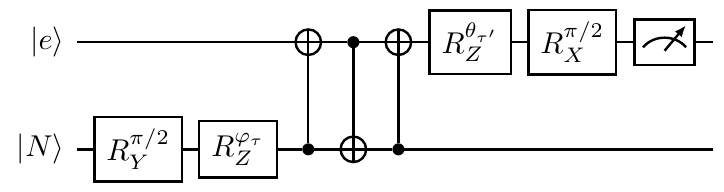}
  \caption{
    An example protocol to cancel the magnetic noise effect.
  }
  \label{fig:comagnetometry}
\end{figure}

Related to the above point, the fact that the NV centers are sensitive to the axion coupling with electrons $g_{aee}$ as well as that with nucleons $g_{ann}$ and $g_{app}$ implies a possible extension of our protocols to mitigate the magnetic noise effect.
Similar to the ideas of comagnetometry~\cite{BrownThesis,KornackThesis,VasilakisThesis,Kornack:2004cs,vasilakis2009limits,Brown:2010dt,Lee:2018vaq,Romalis1} and its application to the axion DM search~\cite{Bloch:2019lcy,Wu:2019exd}, the main goal is to cancel the magnetic noise effect while keeping the axion-induced signal by using the fixed ratio between interaction strengths of the ordinary magnetic field to the electron and nuclear spins.
An example protocol is shown in \cref{fig:comagnetometry}, which is dedicated to canceling the dc-like magnetic noise effect.
The signal obtained by the quantum circuit in \cref{fig:comagnetometry} is determined by the phase difference $\varphi_\tau + \theta_{\tau'}$, where $\varphi_\tau$ ($\theta_{\tau'}$) corresponds to the phase acquired by the free precession of the nuclear-spin (electron-spin) state for the time interval $\tau$ ($\tau'$), respectively.
Since these phases for a small dc magnetic noise $B_{\mathrm{noise}}$ are roughly given by $\varphi_\tau \simeq \gamma_{\ce{^14N}} B_{\mathrm{noise}} \tau$ and $\theta_{\tau'} \simeq \gamma_e B_{\mathrm{noise}} \tau'$, one can in principle make the noise contributions cancel with each other with the choice of $\tau/\tau' \simeq -\gamma_e / \gamma_{\ce{^14N}} > 0$.
A detailed study of the sensitivity of this protocol, including the frequency profile of the magnetic noise, the effect from the overhead time, and the effect from the hyperfine interaction during the free precession, remains as a future project.

When preparing the diamond sample, it is possible to have the majority of the NV centers contain the nitrogen isotope $\ce{^15N}$ by creation via implantation~\cite{Rabeau2006} or by doping during chemical vapor deposition synthesis of diamond.
In this case, we primarily obtain constraints on $g_{app}$ because the nuclear spin of $\ce{^15N}$ is predominantly influenced by proton contributions.
Indeed, the nuclear shell model indicates that the expression of the axion-induced magnetic field \cref{eq:gamma_B_N} is replaced by
\begin{align}
  \gamma_N B_N &\simeq - \frac{1}{3} \frac{g_{app}}{m_p} \sqrt{2\rho_a} v_a,
\end{align}
for $\ce{^15N}$.
Therefore, NV center metrology based on $\ce{^15N}$ spins provides yet another independent piece of information about axion-nucleon couplings, which could also be helpful to distinguish the axion-induced signal from the magnetic noise similar to the idea shown in \cref{fig:comagnetometry}.

Finally, we briefly discuss the current state of experimental NV center sensors in relation to our proposal. The nuclear spin is generally not used for magnetic field sensing, as it is much less sensitive than the electron spin. Therefore, there is limited knowledge of their properties today. Since the nuclear spin coherence times are rather long, readout techniques that require more time become feasible. For example, single-shot readout has the potential to reduce the noise in the system close to the spin-projection noise~\cite{doi:10.1126/science.1189075,Dreau2013}, as investigated in this work. Our work is one step towards a better understanding of the properties of nuclear spin metrology, which will provide us with many future opportunities in the fields of sensing and particle physics.

\begin{acknowledgments}
SC thanks Joseph Bramante for useful discussions.
SC thanks people in the Quantum Engineering Group at MIT for useful discussions and comments on our ideas.
This work was supported by
World Premier International Research Center Initiative (WPI), MEXT, Japan.
This work was also supported by the DOE, Office of Science under contract DE-AC02-05CH11231, partially through Quantum Information Science Enabled Discovery (QuantISED) for High Energy Physics (KA2401032). 
This work was also supported by
JST Moonshot (Grant Number JPMJMS226C), JSPS KAKENHI (Grant Number 20H05661,
23H04390, 20H05661, 24K07010),
and CREST
(JPMJCR23I5), JST.
\end{acknowledgments}

\bibliography{bib}
\bibliographystyle{utphys}

\onecolumngrid
\appendix
\appendixsectionformat

\section{Spin synthesis}
\label{sec:synthesis}

In this appendix, we focus on the synthesis of the spin $S=1/2$ and the orbital angular momentum $L=\ell$ and derive how spin operators act on the eigenstates of the total spin $J$.
Let $\ket{\uparrow}$ and $\ket{\downarrow}$ be the spin-up and spin-down states, respectively, and $\ket{m}$ ($m=-\ell, \dots, \ell$) be the eigenstates of the orbital angular momentum with $L_z = m$.
The synthesized states of these two quantum numbers decompose into two groups with total angular momenta $J=\ell\pm \frac{1}{2}$.
By parametrization of these states as $\ket{J, M}$ with $M=-J,\dots,J$ representing the $z$-component of the total angular momentum, the matrix elements of the total spin operators are characterized by the Clebsch-Gordan coefficients
\begin{align}
  \Braket{J,M\pm 1 | J^{\pm} | J,M} = \sqrt{(J\mp M)(J\pm M + 1)},
\end{align}
with $J^\pm \equiv J^x \pm i J^y$.
Using this expression, we can relate the eigenstates of various spins as follows:
\begin{align}
  \Ket{J=\ell+\frac{1}{2}, M} &= \frac{1}{\sqrt{2J}} \left(
    \sqrt{J+M} \ket{\uparrow} \ket{M-1/2}
    + \sqrt{J-M} \ket{\downarrow} \ket{M+1/2}
  \right), \\
  \Ket{J=\ell-\frac{1}{2}, M} &= \frac{1}{\sqrt{2J+2}} \left(
    \sqrt{J-M+1} \ket{\uparrow} \ket{M-1/2}
    - \sqrt{J+M+1} \ket{\downarrow} \ket{M+1/2}
  \right),
\end{align}
or equivalently,
\begin{align}
  \ket{\uparrow}\ket{m} &= \frac{1}{\sqrt{2\ell+1}} \left(
    \sqrt{\ell+m+1} \Ket{\ell+\frac{1}{2}, m+\frac{1}{2}}
    + \sqrt{\ell-m} \Ket{\ell-\frac{1}{2}, m+\frac{1}{2}}
  \right), \label{eq:synthesis_eq1} \\
  \ket{\downarrow}\ket{m} &= \frac{1}{\sqrt{2\ell+1}} \left(
    \sqrt{\ell-m+1} \Ket{\ell+\frac{1}{2}, m-\frac{1}{2}}
    - \sqrt{\ell+m} \Ket{\ell-\frac{1}{2}, m-\frac{1}{2}}
  \right). \label{eq:synthesis_eq2}
\end{align}
Using \cref{eq:synthesis_eq1,eq:synthesis_eq2}, we can calculate all the non-zero matrix elements of the spin operators as follows:
\begin{align}
  \Braket{J=\ell+\frac{1}{2}, M\pm 1 | S^\pm | J, M} &=
  \frac{\sqrt{(J\mp M)(J\pm M+1)}}{2J}, \\
  \Braket{J=\ell+\frac{1}{2}, M | S^z | J, M} &=
  \frac{M}{2J}, \\
  \Braket{J=\ell-\frac{1}{2}, M\pm 1 | S^\pm | J, M} &=
  -\frac{\sqrt{(J\mp M)(J\pm M+1)}}{2J+2}, \\
  \Braket{J=\ell-\frac{1}{2}, M | S^z | J, M} &=
  -\frac{M}{2J+2},
\end{align}
with $S^{\pm} \equiv S^x \pm i S^y$.
From the above equations, we see that, for a fixed value of $J$, the spin operators effectively act as $SU(2)$ generators in the spin-$J$ representation with a non-trivial factor,
\begin{align}
  \Braket{J,M' | \vec{S} | J,M} = \pm \frac{1}{2\ell+1} \Braket{J,M' | \vec{J} | J,M},
  \label{eq:group_coeff}
\end{align}
for $J=\ell\pm \frac{1}{2}$.

Now the calculation so far can be applied to the case of the $\Ntr$ spin, which has $I=1$ composed of a neutron and a proton in individual $(1p)_{1/2}$ orbitals.
First, each neutron and proton resides in the $1p$ orbital with $\ell=1$, resulting in the total angular momentum $J_\flabel = \ell-1/2 = 1/2$ ($\flabel=n,p$).
According to \cref{eq:group_coeff}, we obtain the effective relationships among operators
\begin{align}
  \vec{S}_\flabel \sim -\frac{1}{3} \vec{J}_\flabel.
  ~~
  (\chi=n,p)
\end{align}
Since either one of the total angular momentum operators, say $\vec{J}_n$, has spin $J_n = 1/2$, we can repeat the same estimation, combining it with $\vec{J}_p$ to obtain $I=1$ states.
Again according to \cref{eq:group_coeff}, the angular momentum operators of nucleons are related to the $\Ntr$ spin operator $\vec{I}$ as
\begin{align}
  \vec{J}_n \sim \vec{J}_p \sim \frac{1}{2} \vec{I}.
\end{align}
Therefore, the coefficient $-1/6$ in \cref{eq:S_I_relation} is successfully reconstructed.

\section{Calculation of the power spectral density and the quantum noise}
\label{sec:PSD}

In this appendix, we provide a detailed calculation of the PSD defined in \cref{eq:PSD_def} starting from \cref{eq:Ctt}.
Firstly, the density matrix of the nuclear spin state in an ensemble of $N$ NV centers before the $j$-th measurement, $\rho_j$, is expressed as
\begin{align}
  \rho_j &\simeq \bigotimes_{\ell=1}^N \rho_{j\ell}, \\
  \rho_{j\ell} &\equiv \frac{1}{2} \left\{
    (1-2\fluorescence_j) \ketbra{+}{+} + \sqrt{1-4\fluorescence_j^2} \ketbra{+}{0} + \sqrt{1-4\fluorescence_j^2} \ketbra{0}{+} + (1+2\fluorescence_j) \ketbra{0}{0}
  \right\},
\end{align}
for fixed values of $\phi$ and $\vec{v}_a$, where $\fluorescence_j$ is given by \cref{eq:S_DC,eq:S_AC,eq:S_DD} for the Ramsey, the Hahn echo, and the dynamic decoupling sequences, respectively, with the replacement $\phi \to m_a t_j + \phi$.
For notational simplicity, we omit the indices $j$ and $\ell$ for each bra and ket, but they are assumed implicitly.
Thus, $\rho_j$ is a $2^N$-dimensional density matrix.

Our next task is to evaluate the trace factor $\MyTr{\rho_{jj'} M_{jj'}^z}$ for various choices of $j$ and $j'$, where $\rho_{jj'} \equiv \rho _j \otimes \rho _{j'}$ and $M_{jj'}^z \equiv M_j^z \otimes M_{j'}^z$ for $j\neq j'$ and $\rho_{jj} \equiv \rho _j$ and $M_{jj}^z \equiv M_j^z M_{j}^z$.
Firstly, when $j=j'$, we obtain
\begin{align}
  \MyTr{\rho_j M_j^z M_j^z}
  &= \frac{1}{4N^2} \sum_\ell 1 + \frac{1}{4N^2} \sum_{\ell\neq \ell'} \MyTr{
    \rho_{j\ell} \otimes \rho_{j\ell'} \sigma_{j\ell}^z \sigma_{j\ell'}^z
  } \notag \\
  &= \frac{1}{4N} + \frac{N(N-1)}{N^2} F_j^2.
\end{align}
For a different time two-point function with $j\neq j'$, we can instead decompose the trace in two parts for the time $t_j$ and $t_{j'}$ and obtain
\begin{align}
  \MyTr{(\rho_j \otimes \rho_{j'}) (M_j^z \otimes M_{j'}^z)}
  &= \MyTr{\rho_j M_j^z}\, \MyTr{\rho_{j'} M_{j'}^z} \notag \\
  &= \fluorescence_j \fluorescence_{j'}.
\end{align}
We can combine these expressions in a compact form
\begin{align}
  \MyTr{\rho_{jj'} M_j^z M_{j'}^z} \simeq \frac{1}{4N} \delta_{jj'} + \fluorescence_j \fluorescence_{j'},
\end{align}
where subleading terms of $N\gg 1$ are neglected.
Recalling that the ensemble average of a single data is zero, $\Braket{M_j} = 0$, the two-point function defined in the main text,
\begin{align}
  C_{jj'} = &\left. \frac{1}{2\pi} \int d\phi\, \frac{1}{4\pi} \int d\hat{v}_a\,
  \frac{1}{2\pi} \int d\phi'\, \frac{1}{4\pi} \int d\hat{v}_a'\,
  \MyTr{
    \rho_{jj'} M_{jj'}^z
  } \right|_{v_a = 10^{-3}} \notag \\
  &\times \left[
    \Theta(|t_j - t_{j'}| - \tau_a) + 8\pi^2 \delta(\phi-\phi') \delta(\hat{v}_a - \hat{v}_a') \Theta(\tau_a - |t_j - t_{j'}|)
  \right],
  \tag{\ref{eq:Ctt}}
\end{align}
is thus calculated as
\begin{align}
  C_{jj'}
  &\simeq \frac{1}{4N} \delta_{jj'} + \frac{1}{2\pi} \int d\phi\, \frac{1}{4\pi} \int d\hat{v}_a\, \fluorescence_j \fluorescence_{j'} \Big|_{\phi'=\phi, \hat{v}_a' = \hat{v}_a, v_a=10^{-3}} \Theta(\tau_a - |t_j - t_{j'}|) \notag \\
  &= \frac{1}{4N} \delta_{jj'} + \mathcal{A} \cos \left[
    m_a (t_j - t_{j'})
  \right] \Theta(\tau_a - |t_j - t_{j'}|),
  \label{eq:Ctt_exp}
\end{align}
with the protocol-dependent coefficient $\mathcal{A}$ defined as
\begin{align}
  \mathcal{A} \equiv \begin{cases}
    \dfrac{\rho_a v_0^2}{27 \tilde{f}_a^2 m_a^2} \sin^2 \dfrac{m_a \tau}{2}, & (\text{Ramsey}) \\[10pt]
    \dfrac{4\rho_a v_0^2}{27 \tilde{f}_a^2 m_a^2} \sin^4 \dfrac{m_a \tau}{4}, & (\text{Hahn echo}) \\[10pt]
    \dfrac{\rho_a v_0^2}{27 \tilde{f}_a^2 m_a^2} \sin^2 \dfrac{m_a \tau}{2} \tan^2 \dfrac{m_a \tau}{2(N_\pi + 1)}, & (\text{DD}) \\[10pt]
  \end{cases}
  \label{eq:A}
\end{align}
where $v_0= 10^{-3}$ denotes the typical axion velocity.

Next, we calculate the PSD $\PSD{k} \equiv \Braket{O_k}$ using the operator $\mathcal{O}_k$ defined as follows:
\begin{align}
  \mathcal{O}_k \equiv \frac{\tau^2}{\tobs} \sum_{j,j'} e^{2\pi i k (j-j')/\Nobs} M_j^z M_{j'}^z.
  \tag{\ref{eq:Ok_def}}
\end{align}
Using a modified expression
\begin{align}
  \PSD{k} = \frac{\tau^2}{\tobs} \sum_{j,j'} e^{2\pi i k (j-j')/\Nobs} C_{jj'},
  \tag{\ref{eq:PSD_def}}
\end{align}
an easy way to accomplish this task is to consider the continuum limit as follows:
\begin{align}
  \PSD{k} &\simeq \frac{1}{\tobs} \int_0^{\tobs} dt\, \int_0^{\tobs} dt'\, e^{i\omega_k (t-t')} C(t, t'),
\end{align}
where $\omega_k \equiv 2\pi k/\tobs$ and the function $C(t,t')$ is defined as a natural extension of $C_{jj'}$ to the continuous choice of time.
By substituting \cref{eq:Ctt_exp} into the above expression, we obtain
\begin{align}
  \PSD{k} \simeq \frac{\tau}{4N} + \begin{cases}
    \dfrac{2\mathcal{A}}{\tobs \Delta\omega_k^2} \sin^2 \dfrac{\tobs \Delta\omega_k}{2}, & (\tobs < \tau_a) \\[10pt]
    \dfrac{2\mathcal{A}}{\tobs \Delta\omega_k^2} \sin^2 \dfrac{\tau_a \Delta\omega_k}{2} + \dfrac{\tobs-\tau_a}{\tobs\Delta\omega_k} \mathcal{A} \sin\left[
      \tau_a \Delta\omega_k
    \right], & (\tobs > \tau_a)
  \end{cases}
  \label{eq:PSD_exp}
\end{align}
with $\Delta\omega_k \equiv \omega_k - m_a$, where we have neglected the fast oscillation terms.

\subsection{Quantum noise on the PSD}

We can evaluate the quantum noise on the PSD without the axion effect as
\begin{align}
  \BPSD{k} \equiv \left.
    \sqrt{\Braket{\mathcal{O}_k^2} - \Braket{\mathcal{O}_k}^2}
  \right|_{B_N = 0}.
  \tag{\ref{eq:dPSD}}
\end{align}
The first term in the square root can be deformed as
\begin{align}
  \Braket{\mathcal{O}_k^2} &= \Braket{ \left(
    \frac{\tau^2}{\tobs} \sum_{j,j'} e^{2\pi i k (j-j')/\Nobs} M_j^z M_{j'}^z
  \right)^2} \notag \\
  &= \Braket{ \left(
    \frac{\tau^2}{\tobs} \sum_{j,j'} e^{2\pi i k (j-j')/\Nobs} \frac{1}{4N^2} \sum_{\ell,\ell'} \sigma_{j\ell}^z \sigma_{j'\ell'}^z
  \right)^2} \notag \\
  &= \frac{\tau^4}{16\tobs[2] N^4} \Braket{ \left(
    \sum_{j,\ell} \bm{1} + \sum_{(j,\ell) \neq (j',\ell')} e^{2\pi i k (j-j')/\Nobs} \sigma_{j\ell}^z \sigma_{j'\ell'}^z
    \right)^2},
\end{align}
where $\bm{1}$ is the identity operator.
To go further, we note that the odd number of Pauli matrices for a certain combination of $(j, \ell)$ leads to the vanishing contribution $\MyTr{\rho_{j\ell} \sigma_{j\ell}^z} \big|_{v_a=0} = 0$.
Thus, the only remaining contribution comes from the terms proportional to the identity matrix.
In the parenthesis of the third line of the previous equation, the first term trivially leads to such a contribution with size $\Nobs[2] N^2$, while the second term also contributes as
\begin{align}
  &\left(
    \sum_{(j,\ell) \neq (j',\ell')} e^{2\pi i k (j-j')/\Nobs} \sigma_{j\ell}^z \sigma_{j'\ell'}^z
  \right)^2 \notag \\
  &= \sum_{(j_1,\ell_1) \neq (j_2,\ell_2)} \sum_{(j_3,\ell_3) \neq (j_4,\ell_4)}  e^{2\pi i k (j_1-j_2+j_3-j_4)/\Nobs} \sigma_{j_1 \ell_1}^z \sigma_{j_2 \ell_2}^z \sigma_{j_3 \ell_3}^z \sigma_{j_4 \ell_4}^z \notag \\
  &= \sum_{(j_1,\ell_1) \neq (j_2,\ell_2)} \left(
    1 + e^{4\pi i k (j_1 - j_2)/\Nobs}
  \right) + \cdots \notag \\
  &= \Nobs N (\Nobs N - 1) + \sum_{j_1,j_2} \sum_{\ell_1,\ell_2} e^{4\pi i k (j_1 - j_2)/\Nobs} - \sum_{j,\ell} 1 + \cdots \notag \\
  &= \Nobs N (\Nobs N - 2) + \Nobs[2] N^2 \delta_{k,0} + \cdots,
\end{align}
where the identity operator $\bm{1}$ is implicit, while dots represent terms with remnant Pauli matrices.
Substituting this result in the original definition, we obtain
\begin{align}
  \BPSD{0} &\simeq \frac{\tau}{2\sqrt{2}N},
  \label{eq:B0_exp} \\
  \BPSD{k\neq 0} &\simeq \frac{\tau}{4N},
  \label{eq:Bk_exp}
\end{align}
where we neglect the subleading terms of $\Nobs$ and $N$.

\subsection{Shot noise on the PSD}
\label{sec:shot_noise}

The quantum state of the NV center is read out by the fluorescence measurement.
The fluctuation of the number of photons detected during the measurement, i.e. the shot noise, affects the PSD, which can be evaluated following the discussion in \cite{taylor2008high,PhysRevA.96.042115}.

Let $\alpha_{0}$ and $\alpha_-$ be the average number of detected photons from the $\ket{S_z = 0}$ and $\ket{S_z = -}$ states, respectively, which are combinations of the emission probability and the collection efficiency of photons.
Both $\alpha_0$ and $\alpha_-$ are increasing functions of the irradiated laser power, which is considered to be fixed in this section.
Then, for an electron density matrix of a single NV center, $\tilde{\rho}_{j\ell}$, the density matrix of the outgoing photon $\rho_{j\ell}^{\mathrm{ph}}$ can be written as
\begin{align}
  \rho_{j\ell}^{\mathrm{ph}} = \Braket{0 | \tilde{\rho}_{j\ell} | 0} \left(
    (1-\alpha_0) \ket{0}_{j\ell}^{\mathrm{ph}} \bra{0} + \alpha_0 \ket{1}_{j\ell}^{\mathrm{ph}} \bra{1}
  \right) + \Braket{- | \tilde{\rho}_{j\ell} | -} \left(
    (1-\alpha_-) \ket{0}_{j\ell}^{\mathrm{ph}} \bra{0} + \alpha_- \ket{1}_{j\ell}^{\mathrm{ph}} \bra{1}
  \right),
\end{align}
where $\ket{0}_{j\ell}^{\mathrm{ph}}$ and $\ket{1}_{j\ell}^{\mathrm{ph}}$ respectively correspond to the final state without and with photon capture.
By remembering that the final CNOT gates in \cref{fig:protocol_DC} and \cref{fig:protocol_AC} maps the nuclear spin state onto the electron spin state, the same quantity can be equivalently expressed in terms of the density matrix $\rho_{j\ell}$ of the nuclear spin as
\begin{align}
  \rho_{j\ell}^{\mathrm{ph}} = \Braket{+ | \rho_{j\ell} | +} \left(
    (1-\alpha_0) \ket{0}_{j\ell}^{\mathrm{ph}} \bra{0} + \alpha_0 \ket{1}_{j\ell}^{\mathrm{ph}} \bra{1}
  \right) + \Braket{0 | \rho_{j\ell} | 0} \left(
    (1-\alpha_-) \ket{0}_{j\ell}^{\mathrm{ph}} \bra{0} + \alpha_- \ket{1}_{j\ell}^{\mathrm{ph}} \bra{1}
  \right).
  \label{eq:rho_ph}
\end{align}
When an ensemble of the NV centers is considered, the expectation number of detected photons is calculated as $\Braket{I_j}^{\mathrm{ph}} = \MyTr{\rho_j^{\mathrm{ph}} I_j}$, where $\rho_j^{\mathrm{ph}} \equiv \bigotimes_\ell \rho_{j\ell}^{\mathrm{ph}}$ and
\begin{align}
  I_j = \sum_\ell \ket{1}_{j\ell}^{\mathrm{ph}} \bra{1}.
\end{align}
Throughout this appendix, for calculational simplicity, we assume that the NV centers in the diamond are prepared to be aligned along one of the four directions of the carbonic covalent bonds using the techniques introduced in \cite{Fukui:2014,Miyazaki:2014,Michl:2014,Lesik:2014}.\footnote{
Without the alignment of the NV centers, only a quarter of the NV centers is sensitive to the magnetic field, while the other three quarters of them induces a large amount of baseline fluorescence, resulting in a larger shot noise.
Overall, it gives rise to the $O(1)$ reduction of sensitivity compared with the setup with aligned NV centers.
}
Neglecting any kind of the inhomogeneity that causes dephasing, we obtain $\Braket{I_j}^{\mathrm{ph}} = N n_j$ with a single expectation number of photons $n_j$ defined as
\begin{align}
  n_j \equiv \Braket{+ | \rho_{j\ell} | +} \alpha_0 + \Braket{0 | \rho_{j\ell} | 0} \alpha_-,
\end{align}
with an arbitrary choice of $\ell$.
Similarly, the fluctuation of the number of photons $\delta I_j$, i.e. the shot noise, can be evaluated as
\begin{align}
  \Braket{\delta I_j \delta I_j}^{\mathrm{ph}} &= \Braket{I_j^2}^{\mathrm{ph}} - \left(
    \Braket{I_j}^\mathrm{ph}
  \right)^2 \\
  &= N^2 n_j (1-n_j),
\end{align}
which is the expected result for the binomial distribution.

In our analysis, we use the set of observables $\left\{ I_j \right\}_j$ to define the PSD, instead of using it directly to extract the magnetic signal.
The corresponding definition of the PSD is given by $\PSD[\mathrm{ph}]{k} \equiv \Braket{O_k^\mathrm{ph}}$ where
\begin{align}
  \mathcal{O}_k^\mathrm{ph} \equiv \frac{\tau^2}{\tobs} \sum_{j,j'} e^{2\pi i k (j-j')/\Nobs} I_j I_{j'},
  \label{eq:Oph}
\end{align}
and the bracket $\Braket{\cdots}$ denotes the ensemble average with the set of density matrices $\left\{ \rho_j^{\mathrm{ph}} \right\}_j$ and the integrals over the axion parameters taking into account the coherence time as in \cref{eq:Ctt}.
Using this notation, we define the variables analogous to $C_{jj'}$ in \cref{eq:Ctt} as
\begin{align}
  C_{jj'}^{\mathrm{ph}} \equiv& \Braket{I_j I_{j'}} \notag \\
  = &\left. \frac{1}{2\pi} \int d\phi\, \frac{1}{4\pi} \int d\hat{v}_a\,
  \frac{1}{2\pi} \int d\phi'\, \frac{1}{4\pi} \int d\hat{v}_a'\,
  \MyTr{
    \rho_{jj'}^{\mathrm{ph}} I_j I_{j'}
  } \right|_{v_a = 10^{-3}} \notag \\
  &\times \left[
    \Theta(|t_j - t_{j'}| - \tau_a) + 8\pi^2 \delta(\phi-\phi') \delta(\hat{v}_a - \hat{v}_a') \Theta(\tau_a - |t_j - t_{j'}|)
  \right],
  \tag{\ref{eq:Ctt}}
\end{align}
with $\rho_{jj'}^{\mathrm{ph}} \equiv \rho_j^{\mathrm{ph}} \otimes \rho _{j'}^{\mathrm{ph}}$ for $j\neq j'$ and $\rho_{jj}^{\mathrm{ph}} \equiv \rho_j^{\mathrm{ph}}$.
They are straightforwardly evaluated as
\begin{align}
  C_{jj'}^{\mathrm{ph}} = N^2 \left(
    \frac{\alpha_0 + \alpha_-}{2}
  \right)^2 + \begin{cases}
    N \left(
      \dfrac{\alpha_0 + \alpha_-}{2}
    \right) - N \left(
      \dfrac{\alpha_0 + \alpha_-}{2}
    \right)^2 + N(N-1) (\alpha_0 - \alpha_-)^2 \mathcal{A}, & (j=j') \\
    N^2 (\alpha_0 - \alpha_-)^2 \mathcal{A} \cos \left[
      m_a (t_j - t_{j'})
    \right] \Theta(\tau_a - |t_j - t_{j'}|). & (j\neq j')
  \end{cases}
\end{align}
Using them, the PSD is expressed, in the continuum limit, as
\begin{align}
  \PSD[\mathrm{ph}]{k} &\simeq \frac{1}{\tobs} \int_0^{\tobs} dt\, \int_0^{\tobs} dt'\, e^{i\omega_k (t-t')} C^{\mathrm{ph}}(t, t').
\end{align}
Note that, compared with the projection noise-limited PSD $\PSD{k}$, $\PSD[\mathrm{ph}]{k}$ has an extra factor in front of the axion effect provided by
\begin{align}
  \frac{d\PSD[\mathrm{ph}]{k}/dB_N^2}{d\PSD{k}/dB_N^2} = 
  \frac{d\PSD[\mathrm{ph}]{k}/d\mathcal{A}}{d\PSD{k}/d\mathcal{A}} \simeq
  N^2 (\alpha_0 - \alpha_-)^2,
  \label{eq:projection-shot-coefficients}
\end{align}
where the final equation is a good approximation when $N \gg 1$.
This observation is important to compare the sensitivity between the cases limited by the projection and shot noises.

For evaluation of the shot noise, it is sufficient to consider the ensemble average without the axion effect, which we denote as $\Braket{\cdots}_0$.
With this convention, some of the important quantities are easily calculated as follows:
\begin{align}
  \Braket{I_j}_0 &= N \navg, \label{eq:f_1p} \\
  \Braket{I_j^2}_0 &= N(N-1) \navg[2] + N \navg, \label{eq:f_2p} \\
  \Braket{I_j^3}_0 &= N(N-1)(N-2) \navg[3] + 3N(N-1) \navg[2] + N \navg, \label{eq:f_3p} \\
  \Braket{I_j^4}_0 &= N(N-1)(N-2)(N-3) \navg[4] + 6N(N-1)(N-2) \navg[3] + 7N(N-1) \navg[2] + N \navg, \label{eq:f_4p}
\end{align}
where $\navg\equiv (\alpha_0 + \alpha_-)/2$ is the average number of detected photons without the axion effect.
Note also that the correlation functions of the different-time operators can be decomposed as, e.g. $\Braket{I_j^2 I_{j'}}_0 = \Braket{I_j^2}_0 \Braket{I_{j'}}_0$, since the shot noises at different times are not correlated with each other.
Although the full expressions are listed, only the highest order terms of $N$ are needed to evaluate the leading contribution to the shot noise.
Using the above expressions, the ensemble average of our observable is evaluated as
\begin{align}
  \Braket{\mathcal{O}_k^{\mathrm{ph}}}_0 &= \frac{\tau^2}{t_{\mathrm{obs}}} \left(
    \sum_j \{ N(N-1) \navg[2] + N \navg \} + \sum_{j\neq j'} e^{2\pi i k (j-j')/\Nobs} (N \navg)^2
  \right) \\
  &= \tau \left(
    \Nobs N^2 \navg[2] \delta_{k,0} + N \navg (1-\navg)
  \right),
\end{align}
where we used $\tobs = \Nobs \tau$.
For the evaluation of the second term, the following identity is used, i.e.
\begin{align}
  \sum_{j\neq j'} e^{2\pi i k (j-j')/\Nobs} = \Nobs (\Nobs \delta_{k,0} - 1).
  \label{eq:pseudo_delta}
\end{align}
Similarly, the ensemble average of the squared observable is evaluated according to the expansion as
\begin{align}
  \Braket{ \left( \mathcal{O}_k^\mathrm{ph} \right)^2 }_0 =
  \Big\langle \Big(
    \frac{\tau^2}{\tobs} &\sum_{j,j'} e^{2\pi i k (j-j')/\Nobs} I_j I_{j'}
  \Big)^2 \Big\rangle_0 \\
  = \frac{\tau^4}{\tobs[2]} \Big(
    &\sum_j \Braket{I_j^4}_0 \notag \\
    &+ \sum_{j\neq j'} 2 \left(
      e^{2\pi i k (j-j')/\Nobs} + e^{2\pi i k (j'-j)/\Nobs}
    \right) \Braket{I_j^3 I_{j'}}_0 \notag \\
    &+ \sum_{j > j'} \left(
      4 + e^{4\pi i k (j-j')/\Nobs} + e^{4\pi i k (j'-j)/\Nobs}
    \right) \Braket{I_j^2 I_{j'}^2}_0 \notag \\
    &+ \sum_{j} \sum_{\substack{j' > j'' \\ j\neq j', j\neq j''}} \Big(
      4e^{2\pi i k (j'-j'')/\Nobs} + 4e^{2\pi i k (j''-j')/\Nobs} \notag \\
      &\qquad\qquad\qquad\qquad + 2e^{2\pi i k (2j-j'-j'')/\Nobs} + 2e^{-2\pi i k (2j-j'-j'')/\Nobs}
    \Big) \Braket{I_j^2 I_{j'} I_{j''}}_0 \notag \\
    &+ \sum_{j > j' > j'' > j'''} \left(
      e^{2\pi i k (j-j'+j''-j''')/\Nobs} + \mathrm{perms.}
    \right) \Braket{I_j I_{j'} I_{j''} I_{j'''}}_0
  \Big),
\end{align}
where the last line contains all the possible permutations of the indices $(j,j',j'',j''')$ in the argument of the exponential.
The summation over indices in each line is evaluated according to the repeated use of the identities analogous to \cref{eq:pseudo_delta}, which results in\footnote{
Precisely speaking, all the following identities except for the second one contains additional terms proportional to $\delta_{k,\Nobs/2}$ when $\Nobs$ is even.
Neglecting these terms is justified if we assume that $\Nobs$ is odd, or simply that $\Nobs \gg 1$ and the probability at which the bin $k=\Nobs/2$ is relevant is negligible.
}
\begin{align}
  \sum_{j > j'} \left(
    e^{4\pi i k (j-j')/\Nobs} + e^{4\pi i k (j'-j)/\Nobs}
  \right) &\simeq \Nobs (\Nobs \delta_{k,0} - 1), \\
  \sum_{j} \sum_{\substack{j' > j'' \\ j\neq j', j\neq j''}} \left(
      e^{2\pi i k (j'-j'')/\Nobs} + e^{2\pi i k (j''-j')/\Nobs}
    \right) &= \Nobs (\Nobs - 2) (\Nobs \delta_{k,0} - 1), \\
  \sum_{j} \sum_{\substack{j' > j'' \\ j\neq j', j\neq j''}} \left(
    e^{2\pi i k (2j-j'-j'')/\Nobs} + e^{-2\pi i k (2j-j'-j'')/\Nobs}
  \right) &\simeq \Nobs \{ \Nobs (\Nobs - 3) \delta_{k,0} + 2 \}, \\
  \sum_{j > j' > j'' > j'''} \left(
    e^{2\pi i k (j-j'+j''-j''')/\Nobs} + \mathrm{perms.}
  \right) &\simeq \Nobs[2] (\Nobs - 3)^2 \delta_{k,0} + 2 \Nobs (\Nobs - 3).
\end{align}
By combining all the terms, we obtain the final expression
\begin{align}
  \Braket{ \left( \mathcal{O}_k^\mathrm{ph} \right)^2 }_0 = \tau^2 \left(
    \Nobs[2] N^4 \navg[4] + 6 \Nobs N^3 \navg[3] (1- \navg) + \cdots
  \right) \delta_{k,0} + \tau^2 \left(
    2 N^2 \navg[2] (1-\navg)^2 + \cdots
  \right),
\end{align}
where the dots represent terms with lower order of $N$ and/or $\Nobs$.
Thus, the fluctuation of the observable is evaluated as
\begin{align}
  \BPSD[\mathrm{ph}]{k} &\equiv \left[
    \Braket{ \left( \mathcal{O}_k^\mathrm{ph} \right)^2 }_0 - \Braket{\mathcal{O}_k^\mathrm{ph}}_0^2
  \right]^{1/2} \\
  &\simeq \tau \sqrt{4 \Nobs N^3 \navg[3](1-\navg) \delta_{k,0} + N^2 \navg[2] (1-\navg)^2}.
\end{align}

Finally, the sensitivities of our approach can be compared between the cases limited by the projection and shot noises.
In terms of the single-bin sensitivity in \cref{eq:single-bin}, the ratio of the sensitivities is calculated as
\begin{align}
  \frac{\BPSD[\mathrm{ph}]{k}}{d\PSD[\mathrm{ph}]{k}/dB_N^2} \left(
    \frac{\BPSD{k}}{d\PSD{k}/dB_N^2}
  \right)^{-1} = \begin{cases}
    \dfrac{\sqrt{2\Nobs N \navg (1-\navg)}}{C^2 \navg}, & (k=0) \\
    \dfrac{(1-\navg)}{C^2 \navg}, & (k\neq 0)
  \end{cases}
  \label{eq:ratio_sigma_R}
\end{align}
where we used \cref{eq:projection-shot-coefficients} and the measurement contrast $C$ defined as~\cite{taylor2008high}
\begin{align}
  C \equiv \frac{\alpha_0 - \alpha_-}{\alpha_0 + \alpha_-}.
\end{align}
For $k\neq 0$ modes, the overall sensitivity is then expressed by the parameter $\sigma_R \equiv \sqrt{1 + 1/(C^2 \navg)}$ when $\navg \ll 1$, or equivalently by the readout fidelity $\mathcal{F} \equiv 1/\sigma_R$.
This results in the sensitivity worse than the projection noise-limited one by a factor of $\sigma_R$, which can be as small as $\sigma_R \simeq 19$~\cite{Barry:2023hon}.
On the other hand, the shot noise for the $k=0$ mode is further enhanced by a large factor of $\sqrt{\Nobs N}$, which arises from the fact that the observable $\mathcal{O}_k^{\mathrm{ph}}$ contains terms linearly affected by the nuclear spin.
As a result, the $k=0$ mode is basically useless in our approach.
This unconventional scaling, which persists even in the limit of the perfect measurement $\alpha_0 = 1$ and $\alpha_- = 0$, is a result of the intrinsic constant shift in the definition of the operators $I_j$, resulting in the finite expectation value $\Braket{I_j}_0 = N \navg$.
In principle, this issue can be addressed by shifting $I_j \to I_j - N\navg$ in the definition of the PSD, \cref{eq:Oph}, with which both the ratios shown in \cref{eq:ratio_sigma_R} become independent on $\Nobs$ and $N$.
In reality, this prescription is highly demanding since calibration of $N\navg$, which is an unknown value a priori, at precision of $1/\sqrt{\Nobs N}$ is required.
However, since the frequency range where this mode has a dominant contribution to the sensitivity is given by $f < \tobs[-1]$, the signal is more efficiently explored by the conventional statistical analysis of the same Ramsey sequence data set, whose sensitivity is again expressed using $\sigma_R$.

Overall, we conclude that the single parameter $\sigma_R$ determines the shot noise-limited sensitivity irrespective of the signal frequency.
We use these observations to plot the shot noise-limited sensitivity in \cref{sec:results}.
Note that $\sigma_R$ could be further reduced, e.g. by working with higher laser power since both $\alpha_0$ and $\alpha_1$ are increasing functions of the laser power.

\end{document}